\documentclass[a4paper,12pt]{article}
\usepackage{amsmath}
\usepackage{amsfonts}
\usepackage{amssymb}
\usepackage{latexsym}
\usepackage{epsfig}
\usepackage{graphicx}
\usepackage{oldgerm}
\usepackage{theorem}
\usepackage{bm}

\usepackage{url}

\usepackage[hang,small,bf]{caption}
\usepackage[subrefformat=parens]{subcaption}
\def\N{\mathbb{N}}
\def\C{\mathbb{C}}
\def\R{\mathbb{R}}
\def\T{\mathbb{T}}
\def\D{\mathbb{D}}
\def\J{\mathbb{J}}

\def\sT{\mathsf{T}}

\def\bra{\langle}
\def\ket{\rangle}

\def\x{\bm{x}}
\def\1{{\bf 1}}

\def\u{\bm{u}}
\def\v{\bm{v}}
\def\w{\bm{w}}
\def\e{\bm{e}}
\def\m{\bm{m}}
\def\c{\bm{c}}

\def\rO{{\rm O}}
\def\rR{{\rm R}}
\def\rI{{\rm I}}

\def\hhat{\widehat{h}}
\def\la{\lambda}
\def\trivial{{\mathbf 1}}

\def\cI{\mathcal{I}}

\theorembodyfont{\itshape}

\newtheorem{thm}{Theorem}[section]
\newtheorem{lem}[thm]{Lemma}
\newtheorem{cor}[thm]{Corollary}
\newtheorem{prop}[thm]{Proposition}

\newtheorem{df}[thm]{Definition}

\newtheorem{rem}[thm]{Remark}
\newtheorem{ex}[thm]{Example}

\newcommand{\SSC}[1]{\section{#1}\setcounter{equation}{0}}
\newcommand{\qed}{\hbox{\rule[-2pt]{3pt}{6pt}}}

\newfont{\bg}{cmr10 scaled\magstep4}
\newcommand{\bigzerol}{\smash{\hbox{\bg 0}}}
\newcommand{\bigzerou}{\smash{\lower1.7ex\hbox{\bg 0}}}

\begin{document}

\title{\bf 
Generalized Eigenspaces and Pseudospectra \\
of Nonnormal and Defective\\ 
Matrix-Valued Dynamical Systems
}
\author{
Saori Morimoto, 
\footnote{
Department of Physics,
Faculty of Science and Engineering,
Chuo University, 
Kasuga, Bunkyo-ku, Tokyo 112-8551, Japan
} \, 
Makoto Katori, 
\footnote{
Department of Physics,
Faculty of Science and Engineering,
Chuo University, 
Kasuga, Bunkyo-ku, Tokyo 112-8551, Japan;
e-mail: 
makoto.katori.mathphys@gmail.com
} \, 
and
Tomoyuki Shirai, 
\footnote{
Institute of Mathematics for Industry, 
Kyushu University, 
744 Motooka, Nishi-ku,
Fukuoka 819-0395, Japan; 
e-mail: shirai@imi.kyushu-u.ac.jp
}
}

\date{11 November 2025}
\pagestyle{plain}
\maketitle

\begin{abstract}
We consider nonnormal matrix-valued dynamical systems
with discrete time.
For an eigenvalue of matrix, 
the number of times it appears as a root of 
the characteristic polynomial is called the 
algebraic multiplicity. 
On the other hand, 
the geometric multiplicity is the dimension
of the linear space of eigenvectors associated
with that eigenvalue.
If the former exceeds the latter, 
then the eigenvalue is said to be defective
and the matrix becomes nondiagonalizable
by any similarity transformation.
The discrete-time of our dynamics is identified with
the geometric multiplicity of the zero eigenvalue
$\lambda_0=0$. 
Its algebraic multiplicity 
takes about half of the matrix size at $t=1$
and increases stepwise in time, which keeps
excess to the geometric multiplicity until 
their coincidence at the final time. 
Our model exhibits relaxation processes from 
far-from-normal to near-normal matrices,
in which the defectivity of $\lambda_0$ is recovering 
in time. We show that such processes are realized as
size reductions of pseudospectrum including $\lambda_0$. 
Here the pseudospectra are the domains
on the complex plane which are not necessarily 
exact spectra but in which the resolvent
of matrix takes extremely large values.
The defective eigenvalue $\lambda_0$ is 
sensitive to perturbation and the eigenvalues of
the perturbed systems are distributed densely in 
the pseudospectrum including $\lambda_0$. 
By constructing generalized eigenspace for
$\lambda_0$, we give the Jordan block
decomposition for the resolvent of matrix
and characterize the pseudospectrum dynamics.
Numerical study of the systems perturbed by
Gaussian random matrices supports the
validity of the present analysis.

\vskip 0.2cm

\noindent{Keywords:} 
Nonnormal matrices,
Defective eigenvalue,
Pseudospectrum dynamics, 
Generalized eigenspaces, 
Jordan decompositions of resolvents, 
Symbol curves

\vskip 0.2cm
\noindent{\it 2020 Mathematics Subject Classification:}
15A18; 15A20; 15B05; 47A10; 60B20

\end{abstract}
\vspace{3mm}


\SSC
{Introduction}
\label{sec:introduction}

Consider  a matrix $M \in \C^{n \times n}$.
We write its Hermitian conjugate as $M^{\dagger}$ 
defined by the complex conjugate 
of the transpose; $M^{\dagger}:=\overline{M^{\sT}}$.
If it satisfies the equality
$M^{\dagger} M = M M^{\dagger}$, then
it is said to be \textit{normal} and 
can be diagonalized by a unitary transformation.
Hermitian matrices satisfying 
$M^{\dagger}=M$ fall in this class.
We introduce two notions of degeneracy of eigenvalues.
The \textit{algebraic multiplicity} is the number of
times the eigenvalue appears as a root of 
the characteristic polynomial of $M$, while
the \textit{geometric multiplicity} is the dimension
of the linear space of the eigenvectors associated
to the eigenvalue.
When the geometric multiplicity is equal to the algebraic multiplicity
for all eigenvalues, the matrix
$M$ has a complete set of eigenvectors.
In this case, even if $M$ is nonnormal, 
it can be reduced to a diagonal matrix $\Lambda$ 
by a similarity transformation as
$\Lambda=V^{-1} M V$. 
Here the $j$-th column of $V$ is given by the 
$j$-th linearly independent eigenvector, $j=1,2, \dots, n$, 
and $V^{-1}$ is well defined.
But if the algebraic multiplicity of an eigenvalue 
exceeds the geometric multiplicity,
then that eigenvalue is said to be \textit{defective} and the matrix
becomes nondiagoralizable. 

Consider the \textit{shift matrix} with size $n \geq 2$, 
\begin{equation}
S=S_n:= \Big(\delta_{j \, k-1} \Big)_{1 \leq j, k \leq n}
=
\left(
\begin{array}{cccccc}
0 & 1 &  &  & \bigzerou & \\
  & 0  & 1 & & & \\
& \ldots & \ldots & & & \\
& & \ldots & \ldots &&  \\
& &  & 0 & 1 &  \\
& &  &   & 0 & 1 \\
& \bigzerol & & & & 0
\end{array}
\right), 
\label{eq:S}
\end{equation}
where $\delta_{j k}$ denotes the Kronecker delta.
It is obvious that $S^k=0$ for $k \geq n$.
Let $b_j \in \C$, $j=1,2,\dots$, and consider
\textit{nilpotent Toeplitz matrices}, 
\begin{equation}
S(m):= S^m+b_1 S^{m+1} + \cdots
+ b_{n-m-1} S^{n-1},
\label{eq:Sm}
\end{equation}
for $m =1,2, \dots, n-1$. 
Notice that $S(m)$, $m=1, 2, \dots, n-1$ are
non-Hermitian and nonnormal.
Since all the elements of the diagonal and 
lower triangular part of $S(m)$ are zero,
\[
\lambda_0 :=0
\]
is the only eigenvalue.
Hence, the algebraic multiplicity of
$\lambda_0$, which is denoted by $a_0(m)$,
is $n$ for any $m=1,2, \dots, n-1$. 
For each $m =1, 2, \dots, n-1$, any vector
$\v_0(m)$ in the form
$\v_0(m)= 
(v_{01}(m), 
v_{02}(m), 
\cdots, 
v_{0 m}(m), 
0, \cdots, 0
)^{\sT}$
can be an eigenvector for $\lambda_0$.
The dimension of the complex space 
spanned by these eigenvectors is $m$. 
Hence the geometric multiplicity $g_0(m)$ 
of $\lambda_0$ is $m$. 
We notice that 
\[
g_0(m) < a_0(m), \quad m=1, 2, \dots, n-1.
\]
That is, the eigenvectors fail to span $\C^n$.
In this case, the eigenvalue $\lambda_0$ is defective and
there is no similarity transformation 
which reduces $S(m)$ to any diagonal form.

Although $S(m)$ is nondiagonalizable, 
a similarity transformation can reduce the matrix
to the \textit{Jordan canonical form},
\[
\J(m)=V(m)^{-1} S(m) V(m).
\]
The diagonal elements of $\J(m)$ are still all zero,
but some of the upper 2nd-diagonal elements can be 1.
More precisely, $\J(m)$ is decomposed into
$g_0(m)$-ple \textit{Jordan blocks}
which are given by shift matrices \eqref{eq:S},
\begin{equation}
\J(m)=\bigoplus_{\ell=1}^{g_0(m)} S_{d_0^{(\ell)}(m)},
\label{eq:J0}
\end{equation}
where the sizes of them are denoted by
$d_0^{(\ell)}(m)$, $\ell=1, 2, \dots, g_0(m)$.
By convention, we assume the non-increasing order, 
$d_0^{(1)}(m) \geq d_0^{(2)}(m) \geq 
\cdots \geq d_0^{(g_0(m))}(m)$.
Each $d_0^{(\ell)}(m)$ gives the dimension
of the \textit{generalized sub-eigenspace} which includes
the $\ell$-th eigenvector $\v_0^{(\ell)}(m)$ 
associated with the zero-eigenvalue $\lambda_0$,
$\ell=1,2, \dots, g_0(m)$.
They satisfy the
sum rule $\sum_{\ell=1}^{g_0(m)} d_0^{(\ell)}(m)=a_0(m)$.
By definition, if $d_0^{(\ell)} \equiv 1$, $\ell=1,2, \dots, g_0$,
then $g_0=a_0$ and the matrix is normal.
Therefore,
\begin{equation}
k_0(m) := \max\{ d_0^{(\ell)}(m):  \ell=1,2, \dots, g_0(m)\}
=d_0^{(1)}(m)
\label{eq:k0}
\end{equation}
will indicate the degree of defectivity of
the zero-eigenvalue $\lambda_0$ measuring 
how far from being normal.
It is called the \textit{index} of $\lambda_0$.

Eigenvalue analysis is one of the most successful method
of applied mathematics.
Roughly speaking, eigenvalues give us the abstraction
of a matrix by plotting the set of them
(\textit{spectra}) on a complex plane $\C$.
For defective matrices, however, eigenvalue analysis fails.
In the above case with $S(m)$,
the unique eigenvalue at zero with
full algebraic multiplicity $a_0(m) \equiv n$ 
misleads us to identify $S(m)$ with the
null (all-zero) matrix $O$. 
The Jordan-block structure with zero diagonal
can not be represented only by spectra.

\begin{figure}[htbp]
    \begin{tabular}{ccc}
    \hskip -2.5cm
      \begin{minipage}[t]{0.63\hsize}
       \centering
        \includegraphics[keepaspectratio, scale=0.25]{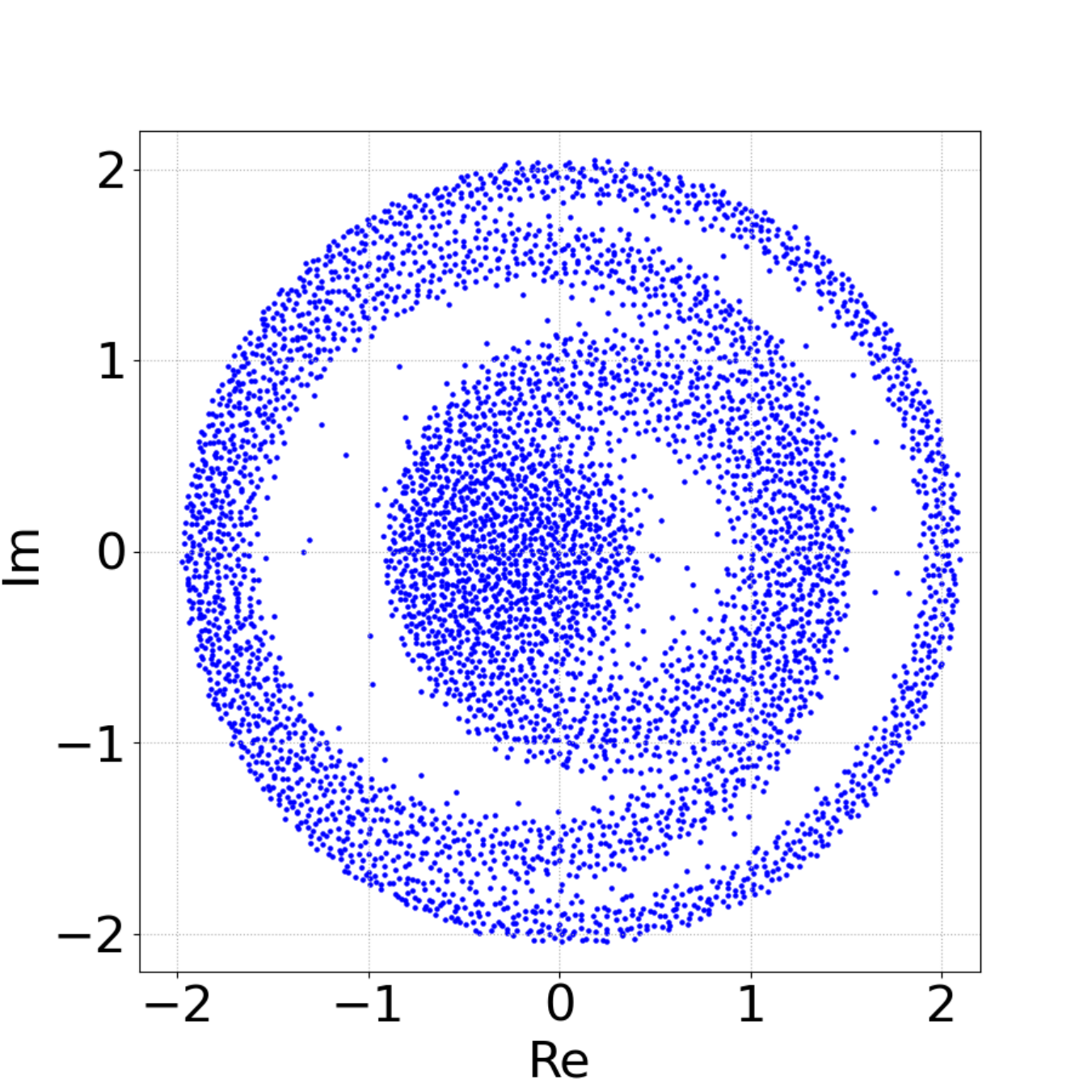}
        \subcaption{}
        \label{fig:SdZa}
      \end{minipage} &
    \hskip -5.3cm
      \begin{minipage}[t]{0.63\hsize}
       \centering
        \includegraphics[keepaspectratio, scale=0.25]{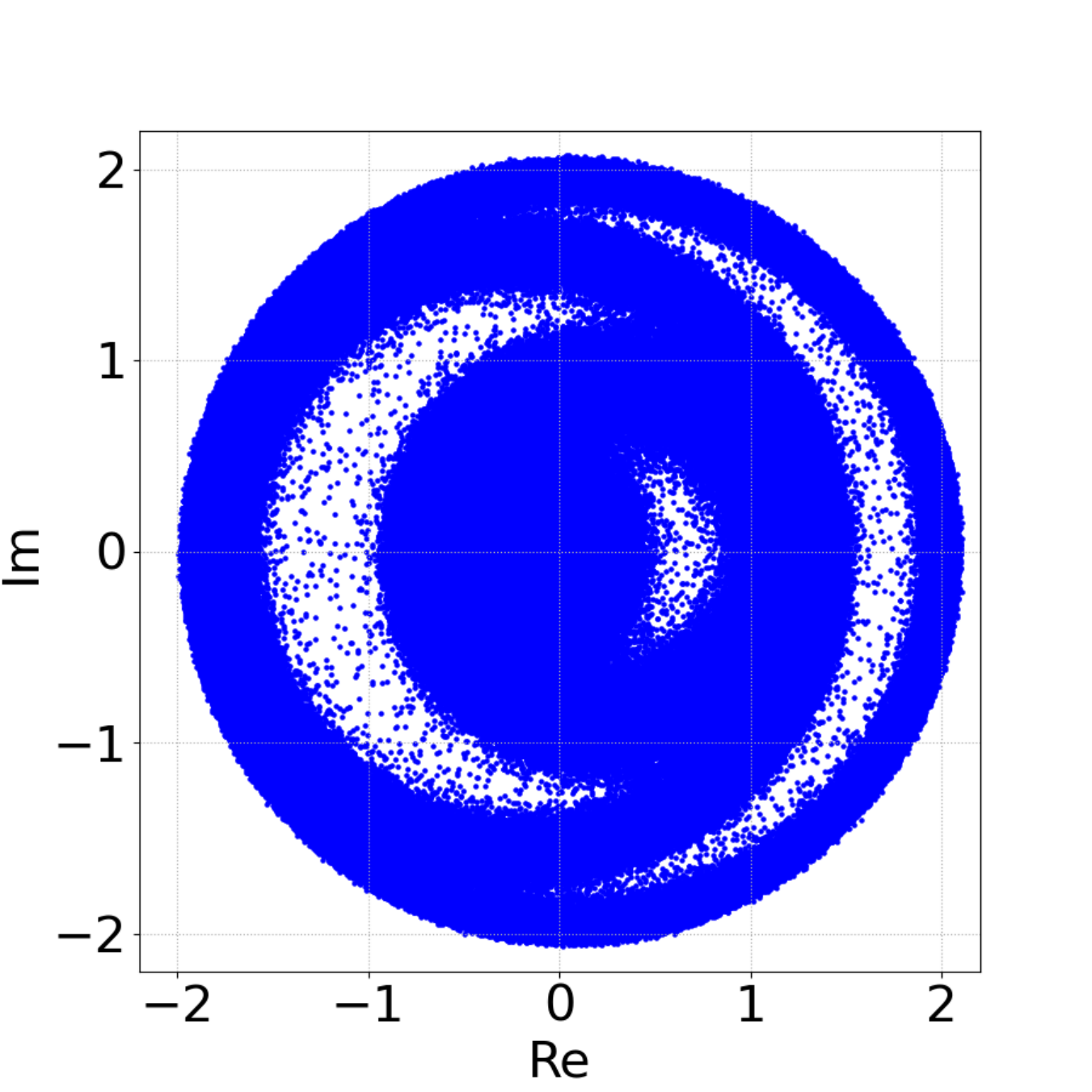}
        \subcaption{}
        \label{fig:SdZb}
      \end{minipage} &
          \hskip -5.3cm
      \begin{minipage}[t]{0.63\hsize}
       \centering
        \includegraphics[keepaspectratio, scale=0.25]{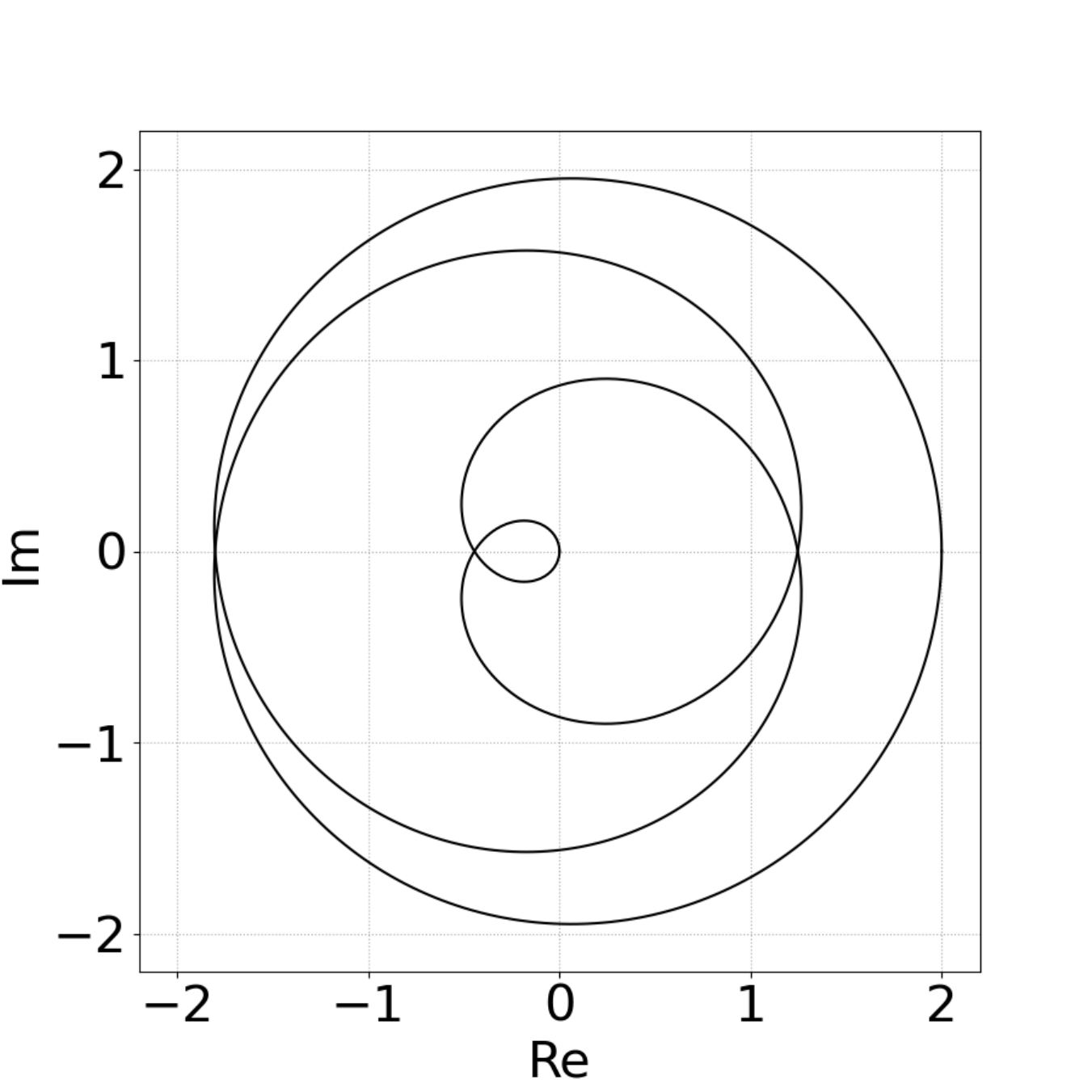}
        \subcaption{}
        \label{fig:SdZc}
      \end{minipage} 
    \end{tabular}
     \caption{(a) One sampling result of numerically obtained 
     eigenvalues of the system  with Gaussian perturbation matrix $Z$;
     $S^{(b)}(m, \delta Z)=S^m+bS^{m+1}+ (1/\sqrt{2n}) Z$, 
     where
     $n=5000$, $m=3$, and $b=1$.
     (b) Superpositions of 50 samples.
     (c) Symbol curve of 
     $\widehat{S}^m+ b \widehat{S}^{m+1}$ with
     $m=3$ and $b=1$.}
     \label{fig:SdZ}
  \end{figure}

The structure of $S(m)$ hidden in eigenvalue analysis
is revealed by its sensitivity to perturbation.
Let $Z=(Z_{jk})_{1 \leq j, k \leq n}$ 
be a matrix consisting of independently and normally
distributed complex entries;
\begin{equation}
Z_{jk} =X_{jk}+i Y_{jk} \quad
\mbox{with} \quad
X_{jk} \sim {\rm N}(0, 1), \quad
Y_{jk} \sim {\rm N}(0, 1), 
\label{eq:Z}
\end{equation}
where $i :=\sqrt{-1}$.
Here we consider a simple case for \eqref{eq:Sm} such that
$b_1=b \in \C$, $b_j=0$ for $j \geq 2$. 
Figure \ref{fig:SdZa} shows plots of the
numerically obtained eigenvalues, 
when the complex Gaussian random matrix $Z$ is added, 
\begin{equation}
S^{(b)}(m, \widetilde{\delta} Z) 
:=S^m+b S^{m+1}+ \widetilde{\delta} Z, 
\label{eq:S_Gauss}
\end{equation}
where $n=5000$, $m=3$, $b=1$, 
and the coefficient of $Z$ is given by
$\widetilde{\delta} =\widetilde{\delta}(n)
=1/\sqrt{2n}=1/100$.
The dots are not the eigenvalues of the original matrix, 
$S^{(b)}(m, 0)=S^m+b S^{m+1}$, but
they represent the structure of
this defective matrix in the sense explained below. 
Hence they are called the \textit{pseudo-eigenvalues}
of $S^{(b)}(m, 0)$. 
(More precisely speaking, the domains on $\C$ in which
the eigenvalues of randomly perturbed matrix are 
distributed are called the \textit{pseudospectra}
of the original unperturbed matrix 
having defective eigenvalues \cite{BS99,RT92,TE05}.)
Instead of the $n \times n$ shift matrix \eqref{eq:S},
here we consider an infinite-size shift matrix,
\begin{equation}
\widehat{S}:=S_{\infty}
=(\delta_{j \, k-1})_{1 \leq j, k < \infty}.
\label{eq:Shat}
\end{equation}
which will represent a \textit{nilpotent Toeplitz operator}.
We also consider an infinite-dimensional 
vector $\widehat{\v}$ in the form
$\widehat{\v}=(1, z, z^2, z^3, \cdots )^{\sT}$, $z \in \C$.
Then we see that
\begin{equation}
(\widehat{S}^m+b \widehat{S}^{m+1}) \widehat{\v}
=f(z) \widehat{\v}
\quad
\mbox{with} \quad
f(z)=z^m+ b z^{m+1}.
\label{eq:Shat2}
\end{equation}
The 2-norm of $\widehat{\v}$,
$\| \widehat{\v} \|
:=(\sum_{j=1}^{\infty} |\widehat{v}_j|^2 )^{1/2}
=(1-|z|^2)^{-1/2}$, 
is finite if $z$ is inside the unit circle
$\T:=\{z \in \C: |z|=1\}$.
The function $f(z)$ is known as the
\textit{symbol} of the Toeplitz operator
$\widehat{S}^m+b \widehat{S}^{m+1}$.
The image of $\T$ by the map $f$, $f(\T)$, is called
the \textit{symbol curve} 
of $\widehat{S}^m+b \widehat{S}^{m+1}$
and is drawn in Fig.\ref{fig:SdZc} for $b=1$ \cite{BS99,TE05}.
In Fig.\ref{fig:SdZa}, the plots of eigenvalues
of the perturbed systems 
line up along the symbol curve $f(\T)$.
For all $z \in \C$, $|z| < 1$, 
the vectors $\widehat{\v}$ satisfying \eqref{eq:Shat2}
give eigenvectors of eigenvalues 
$f(z)=z^m+b z^{m+1}$, $m-1,2, \dots, n-1$ in $\ell^2$-space.
It is proved that the spectrum of 
the Toeplitz operator $\widehat{S}^m+b \widehat{S}^{m+1}$
is equal to $f(\T)$ with all the points enclosed by this 
curve (see, \cite[Theorem 2.1]{RT92}, 
\cite[Theorem 7.1]{TE05}),
and Section \ref{sec:future} (i) below). 
If numerically obtained eigenvalues of \eqref{eq:S_Gauss} 
were all superposed, they will fulfill the region 
bounded by the outmost curve of the
symbol curve $f(\T)$. 
Figure \ref{fig:SdZb} shows superposition
of the plots of 50 samples.
Mathematical study of random perturbations of 
banded Toeplitz matrices, see 
\cite{BGKS22,BKMS21,BPZ19,BPZ20,BZ20,SV21} 
and references therein. 

Recently, the notion of pseudospectra has attracted
much attention in stochastic analysis of time-dependent
random matrix theory \cite{For10,Kat16}.
The non-Hermitian matrix-valued Brownian motion (BM)
is defined by 
\[
M(t)=(M_{jk}(t))_{1 \leq j, k \leq n}
:= \left( \frac{1}{\sqrt{2n}}(
B^{\rR}_{jk}(t)+ i B^{\rI}_{jk}(t)) \right)_{1 \leq j, k \leq n},
\quad t \geq 0, 
\]
where $(B^{\rR}_{jk}(t))_{t \geq 0}$,
$(B^{\rI}_{jk}(t))_{t \geq 0}$, $1 \leq j, k \leq n$
are $2n^2$ independent one-dimensional standard BMs
\cite{BCH24,BD20,Burda15,EKY23,Fyo18,GW18}.
This process is regarded as the dynamical extension
of the complex Ginibre ensemble, 
since if it starts from the null matrix, $M(0)=O$,
the distribution of eigenvalues at an arbitrary time $t > 0$ is 
identified with the 
\textit{complex Ginibre ensemble} of eigenvalues
with variance $t$ \cite{Gin65}, which has been 
extensively studied 
in random matrix theory \cite{BF24,For10}.
Burda et al.~\cite{Burda15} studied the process 
$(M(t))_{t \geq 0}$ starting from
$M(0)=S$. 
By numerical simulation they found that 
the eigenvalues seem to expand instantly from
$\lambda_0$ to a unit circle $\T$.
For the time interval $0 < t < 1$,
the dots form a growing annulus. 
Then the inner radius of the annulus shrinks to
zero at $t=1$ and dots fill up a full unit disk.
This observation shall be compared with
the eigenvalues
of the matrix \eqref{eq:S_Gauss}
with $m=1$ and $b=0$, where perturbation 
was given by a Gaussian complex random-matrix $Z$.
When $m=1$ and $b=0$, the symbol is $f(z)=z$ and hence
the symbol curve is simply a unit circle, $f(\T)=\T$.
The transition from a unit circle to a unit disk
in the time period $t \in (0, 1]$ 
reported by Burda et al.~\cite{Burda15} will be
regarded as a sampling process of eigenvalues
for the perturbed nilpotent Toeplitz matrix.
They seem to cluster along the symbol curve $\T$ when the
sampling number is small, while they tend
to fill the unit disk as the sampling number becomes
large. 

Motivated by the above consideration, 
we will study the following 
matrix-valued dynamical system \cite{MKS24}, 
\begin{equation}
S^{(b)}(t, \delta J) 
:= S^{t+1} + b S^{t+2}+\delta J,
\label{eq:model}
\end{equation}
with discrete time $t=0, 1, \dots, T$,
where the final time is defined by 
\[
T=T(n):=n-2.
\]
Here $b, \delta \in \C$ and 
$J$ is the all-ones matrix; 
$J=(J_{jk})_{1 \leq j, k \leq n}$ with
\[
J_{jk} \equiv 1, \quad j, k =1,2, \dots n, 
\]
which gives the additive 
\textit{rank 1 perturbation} \cite{For23}
to the nilpotent Toeplitz matrices,
$S^{(b)}(t, 0)=S^{t+1}+b S^{t+2}$. 

\begin{figure}[ht]
\begin{center}
\includegraphics[width=0.7\textwidth]{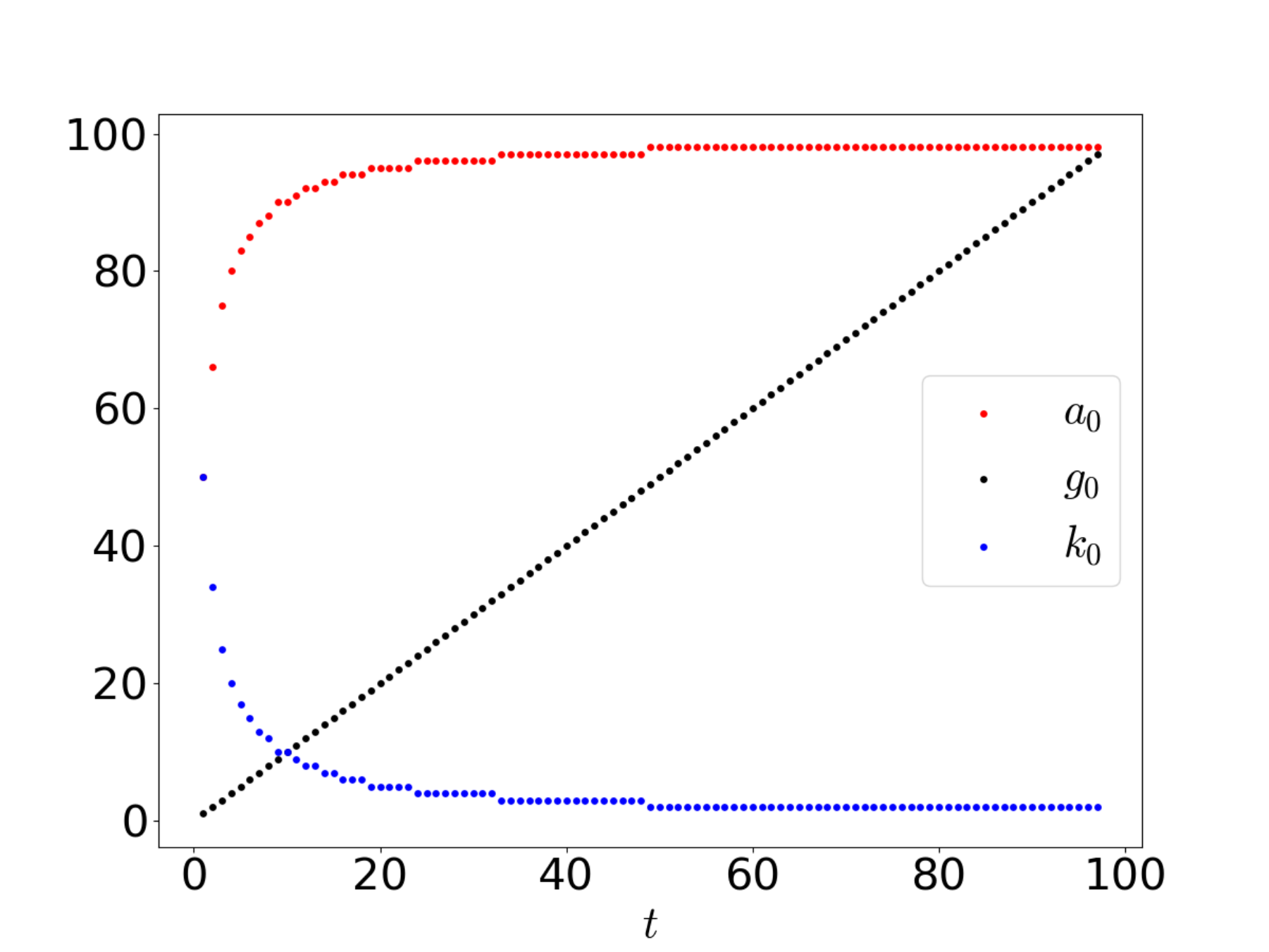}
\end{center}
\caption{
Time-evolution of the geometric multiplicity $g_0$, 
the algebraic multiplicity $a_0$, and 
the degree $k_0$ of $\lambda_0$ 
are shown by black, red, and blue dots, respectively,
for $t=1, 2, \dots, T:=n-2$ with $n=100$.}
\label{fig:relax}
\end{figure}
\begin{figure}[ht]
\begin{center}
\includegraphics[width=1.0\textwidth]{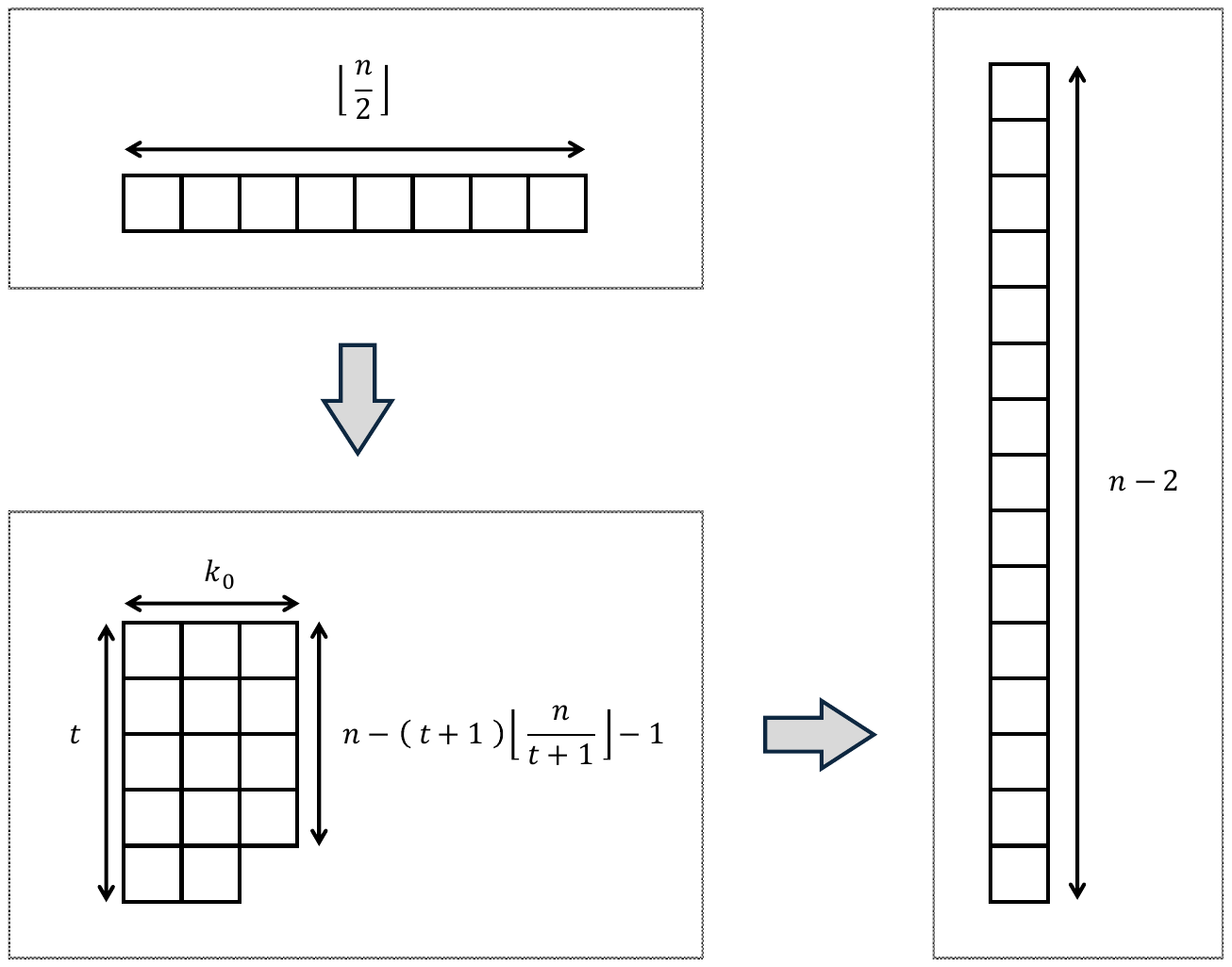}
\end{center}
\vskip -1cm
\caption{
Time evolution of Young diagram representing
the dynamics of the Jordan canonical form
associated with $\lambda_0$.
The upper-left diagram is for the initial time $t=1$ 
with $\lfloor n/2 \rfloor$ boxes and
the right diagram is for the final time $t=T$
with $n-2$ boxes.
Here we have drawn the case with $n=17$. 
The lower-left diagram shows the intermediate state
at time $t=5$.
In this case, $\xi(t,n):=n\bmod{(t+1)}=5$.
Then $k_0=\lfloor n/(t+1) \rfloor +1=3$ and
$n-(t+1) \lfloor n/(t+1) \rfloor -1=\xi(t,n)-1=4$.
}
\label{fig:Young}
\end{figure}

We will show the following. 
\begin{description}
\item{(i)} \quad
The discrete time $t$ is equal to the 
geometric multiplicity of $\lambda_0$, 
\begin{equation}
g_0(t)=t, \quad t=0,1, \dots, T.
\label{eq:g0}
\end{equation}
(See Proposition \ref{thm:g0} in Section \ref{sec:generalES}.)
In other words, we consider the 
matrix-valued dynamical system
such that the dimension of the eigenspace in 
the narrow sense associated with $\lambda_0$ 
is growing in time.

\item{(ii)} \,
All non-zero eigenvalues of $S^{(b)}(t, \delta J)$
are given by simple roots of
a polynomial equation with degree $p_1(t)+1$ with
\begin{equation}
p_1(t) := \left\lfloor \frac{n-1}{t+1} \right\rfloor,
\label{eq:p1}
\end{equation}
where $\lfloor x \rfloor$ denotes the greatest integer less 
than or equal to $x \in \R$ (the floor function of $x$).
This determines the algebraic multiplicity of $\lambda_0$ as
\begin{equation}
a_0(t)=n- p_1(t)-1, \quad t=0,1, \dots, T.
\label{eq:a0}
\end{equation}
(See Theorem \ref{thm:ev1} in Section \ref{sec:eigenvalue}.)
Note that with a fixed $n$, $a_0(t)$ is increasing stepwise
in $t$.
Notice that $a_0(0)=0$; that is, the origin is 
not an eigenvalue at $t=0$. 
When the origin becomes an eigenvalue
at $t=1$, its algebraic multiplicity has
a large value $\lfloor n/2 \rfloor$,
that is, approximately half of the matrix size $n$, 
while the system has only one ($g_0(1)=1$) eigenvector.
Non-zero eigenvalues are being absorbed one by one
to the origin \cite{MKS24}, and the algebraic multiplicity
$a_0(t)$ increases stepwise in time.
The speed of increment of $a_0(t)$ is
slowing down as $t$ increases, but it keeps
excess to $g_0(t)$ until $t=n-3$.
At the final time $t=T$, $\lambda_0$ 
has $a_0(T)=n-2=T$
leaving only two non-zero eigenvalues on $\C$.
There are $g_0(T)=T$ linearly independent
eigenvectors at that time. The coincidence 
$a_0(T)=g_0(T)$ makes the matrix be diagonalizable. 

\item{(iii)}
Although $g_0(0)=a_0(0)=0$ and $g_0(T)=a_0(T)=T$,
the inequality holds as
\[
g_0(t) < a_0(t),
\quad t=1, 2, \dots, T-1.
\]
That is, the matrices $S^{(b)}(t, \delta J)$ are
nonnormal and $\lambda_0$ is defective.
Let
\begin{equation}
\xi(t,n):=n\bmod{(t+1)}, 
\label{eq:xi}
\end{equation}
where $\xi(t,n) \in \{0,1,\dots,t\}$. 
We will construct the generalized eigenspaces
associated with $\lambda_0$ 
explicitly and prove that the 
\textit{index} of $\lambda_0$ is given by
\begin{equation}
k_0(t)= 
\begin{cases}
\displaystyle{
\left\lfloor \frac{n}{t+1} \right\rfloor},
&\quad 
\mbox{if $\xi(t,n) \in \{0,1\}$},
\cr
\cr
\displaystyle{\left\lfloor \frac{n}{t+1} \right\rfloor+1},
&\quad 
\mbox{if $\xi(t,n) \in \{2,3,\dots,t\}$}.
\end{cases}
\label{eq:k02}
\end{equation}
(See Proposition \ref{thm:k0} in Section 
\ref{sec:generalES}.)
The degree of nonnormality is the highest value
at the beginning; $k_0(1) =n/2$ if $n$ is even
and $k_0(1)=(n+1)/2$ if $n$ is odd.
The defectivity of $\lambda_0$ 
is then decreasing stepwise in time.
We put $k_0(T)=1$ by convention. 
See Fig.\ref{fig:relax}

\item{(iv)} 
At each time $t=1,2, \dots, T-1$, 
by a similarity transformation,
$S^{(b)}(t, \delta J)$ is reduced to the following
form, 
\begin{equation}
\J(t) \oplus {\rm diag}(\lambda_1(t), \lambda_2(t),
\dots, \lambda_{p_1(t)+1}),
\label{eq:JordanA1}
\end{equation}
where the Jordan canonical form $\J(t)$
is decomposed into $t$-ple Jordan blocks as
\begin{equation}
\J(t) =
\begin{cases}
\underbrace{S_{k_0(t)} \oplus \cdots \oplus S_{k_0(t)}}_{t},
& \mbox{if $\xi(t,n) \in \{0,1\}$},
\cr
\underbrace{S_{k_0(t)} \oplus \cdots 
\oplus S_{k_0(t)}}_{\xi(t,n)-1}
\oplus
\underbrace{S_{k_0(t)-1} \oplus \cdots 
\oplus S_{k_0(t)-1}}_{t-\xi(t,n)+1},
& \mbox{if $\xi(t,n) \in \{2,3,\dots,t\}$}.
\end{cases}
\label{eq:JordanA2}
\end{equation}
(See Section \ref{sec:generalES}.)
The Jordan canonical form is represented
by a \textit{Young diagram} \cite{CM93,Gan81}.
Figure \ref{fig:Young} shows dynamics of Young diagram
corresponding to \eqref{eq:JordanA2}.
It starts from a single row diagram with 
$\lfloor n/2 \rfloor$ boxes at time $t=1$ and
ends with a single column diagram with $n-2$ boxes
at time $t=T$.
(As mentioned at the end of (ii) above, 
at the final time $t=T$, $S^{(b)}(T, \delta J)=S^{n-1}+\delta J$
can be diagonalized into the form, 
${\rm diag}(\underbrace{0, \cdots, 0}_{T}, \lambda_1(T), \lambda_2(T))$, 
$|\lambda_1(T)| > |\lambda_2(T)| > 0$
instead of \eqref{eq:JordanA1}.) 
The time $t$ is identified with the number of rows
and the total number of boxes is given by
$a_0(t)$ which increases stepwise in time.
The relaxation process of the defectivity of $\lambda_0$
is represented by the stepwise decreasing of the
number of columns.
\end{description}

In summary, the present dynamical systems
$(S^{(b)}(t, \delta J))_{1 \leq t \leq T}$, 
$b, \delta \in \C$,  
represent relaxation processes of matrices
from being far-from-normal to being near-normal, 
in which
the defective eigenvalue $\lambda_0$ 
is recovering in time. 
The paper is organized as follows.
In Section \ref{sec:eigenvalue}
we derive the polynomial equation
for the non-zero eigenvalues and
discuss their time-evolution.
Section \ref{sec:PS} is devoted to
the analytical study of the pseudospectra,
in particular, the pseudospectrum including
the defective eigenvalue $\lambda_0=0$.
In Section \ref{sec:generalES} we explain
how to construct the generalized eigenspace
associated with $\lambda_0$, which is
decomposed into $t$ sub-spaces
and gives the $t$-ple Jordan blocks 
for the similarity transformation of
the matrix at each time $t$, 
We perform the Jordan decomposition
of resolvent of the matrix, introduce the
notion of generalized condition number, 
and then evaluate the 2-norm of
the resolvent (Proposition \ref{thm:resolvent1})
in Section \ref{sec:Jordan}.
In Section \ref{sec:Pseudo}, we give
the first definition of 
$\varepsilon$-pseudospectrum
(Definition \ref{thm:PS}).
Then an estimation of the $\varepsilon$-pseudospectra
is given by Theorem \ref{thm:upperPS}.
This theorem clarifies the
$(t, n)$-dependence of the pseudospectrum
including $\lambda_0$.
In order to show the validity of our analytical
study, we perform numerical calculation
for the processes perturbed by
Gaussian complex random matrices in
Section \ref{sec:numerical}.
In Section \ref{sec:sensitive},
we show the second definition of
$\varepsilon$-pseudospectrum 
(Definition \ref{thm:PS2}) which is
equivalent with the first one.
By this definition, the pseudospectrum
is characterized by the sensitivity of
$\lambda_0$ with respect to perturbations.
Actually, we show in Section \ref{sec:t_n}, that
$(t, n)$-dependence of the
domain, in which the eigenvalues of the perturbed systems
are accumulated, is consistent with 
Theorem \ref{thm:upperPS} for
the pseudospectrum including $\lambda_0$.
Finally in Section \ref{sec:future}
future problems are listed out. 

\SSC
{Eigenvalue Dynamics}
\label{sec:eigenvalue}

For the matrix-valued dynamical system
$(S^{(b)}(t, \delta J))_{0 \leq t \leq T}$ 
given by \eqref{eq:model}, 
we consider the
following eigenvalue and right-eigenvector equations, 
\begin{equation}
S^{(b)}(t, \delta J) \v(t) = \lambda(t) \v(t),
\quad t=0, 1, \dots, T,
\label{eq:ev1}
\end{equation}
$\v(t)=(v_1(t), \cdots, v_n(t))^{\sT} \in \C^n$.
Let $\1$ be the all-ones vector
and we introduce the Hermitian inner product,
\begin{equation}
\bra \u, \v \ket:= \sum_{j=1}^n u_j \overline{v_j}
=\u^{\sT} \overline{\v},
\quad
\u, \v \in \C^n. 
\label{eq:inner}
\end{equation}
Define
\begin{equation}
\alpha(\v)
:= \bra \v, \1 \ket =\sum_{j=1}^n v_j.
\label{eq:alpha}
\end{equation}
By definition, we have the equality
\begin{equation}
J \v = \alpha(\v) \1. 
\label{eq:Jv=a1}
\end{equation}
We set $z=\lambda(t)$.
Then \eqref{eq:ev1} is written as
\begin{equation}
\{ zI-(S^{t+1}+ b S^{t+2} ) \} \v(t)
=\delta \alpha(\v(t)) \1.
\label{eq:ev1b}
\end{equation}
If we consider the zero-eigenvalue $z=\lambda(t)=0$,
\eqref{eq:ev1b} becomes 
$-(S^{t+1}+b S^{t+2}) \v(t)=\delta \alpha(\v(t)) \1$.
Since $S^{t+1}$ and $S^{t+2}$ shift the elements 
of any vector upward by $t+1$ and $t+2$ respectively
when $S^{t+1}+b S^{t+2}$ is operated on the vector from the left,
the last $t+1$ elements of the vector 
$-(S^{t+1}+b S^{t+2}) \v(t)$ are zero. 
Since $\1$ is the all-ones vector, 
$\alpha(\v(t))$ should be 0.
For non-zero eigenvalues $z=\lambda(t) \not=0$ 
on the other hand, 
we can assume $\alpha(\v(t)) \not=0$. 

Let $1_{(\omega)}$ be the indicator function
of the condition $\omega$;
$1_{(\omega)}=1$ if $\omega$ is satisfied,
and $1_{(\omega)}=0$ otherwise.
Remember that $p_1(t)$ is given by \eqref{eq:p1} and
put
\begin{equation}
p_2(t) := \left\lfloor \frac{n-1}{t+2} \right\rfloor.
\label{eq:p2}
\end{equation}

\begin{thm}
\label{thm:ev1}
For $t =0, 1, \dots, T$, the following holds.
\begin{description}
\item{\rm (i)} \,
There are $p_1(t)+1$ non-zero eigenvalues, which solve 
equation, 
\begin{align}
&\frac{1+b}{n \delta} 
\left(\frac{z}{1+b} \right)^{p_1(t)+1}
- \frac{ \displaystyle{1-\left(\frac{z}{1+b} \right)^{p_1(t)+1}} }
{\displaystyle{1-\frac{z}{1+b} } }
\nonumber\\
& \,
+ \left\{
\frac{t+1}{n} + \frac{b}{(1+b)n}
\right\}
\frac{1}{\displaystyle{1-\frac{z}{1+b}}}
\left\{ p_1(t)+1
- \frac{ \displaystyle{1-\left(\frac{z}{1+b} \right)^{p_1(t)+1}} }
{\displaystyle{1-\frac{z}{1+b} } }
\right\}
\nonumber\\
& \,
- \frac{1_{(p_1(t) \geq p_2(t)+1, \, p_1(t) \geq (n+1)/(t+2))}}{(1+b)^{p_1(t)}}
\sum_{k=0}^{p_1(t)-p_2(t)-1} z^k
\sum_{q=n-(t+1)(p_1(t)-k)+1}^{p_1(t)-k} b^q
\binom{p_1(t)-k}{q}
\nonumber\\
& \hskip 6.5cm \times
\frac{1}{n} [ q-\{n-(t+1)(p_1(t)-k)\}]
=0.
\label{eq:eigenvaluesG1}
\end{align}
This equation is also written as the polynomial equation, 
\begin{align}
&
\left( \frac{z}{1+b} \right)^{p_1(t)+1}
- \frac{n \delta}{1+b}
\sum_{k=0}^{p_1(t)} \left[
1-  (p_1(t)-k) 
\left\{
\frac{t+1}{n} + \frac{b}{(1+b) n}
\right\}
\right]
\left(\frac{z}{1+b} \right)^k
\nonumber\\
& \quad
- 1_{(p_1(t) \geq p_2(t)+1, \, p_1(t) 
\geq (n+1)/(t+2))} \frac{n \delta}{(1+b)^{p_1(t)+1}}
\nonumber\\
& \quad \times
\sum_{k=0}^{p_1(t)-p_2(t)-1} z^k
\sum_{q=n-t(p_1(t)-k)+1}^{p_1(t)-k} b^q
\binom{p_1(t)-k}{q}
\frac{1}{n} [ q-\{n-(t+1)(p_1(t)-k)\}]
=0.
\label{eq:eigenvalues}
\end{align}
The corresponding right-eigenvectors $\v(t)$
satisfy $\alpha(\v(t)) \not=0$. 
\item{\rm (ii)} \,
All other $n-p_1(t)-1$ eigenvalues degenerate at zero.
That is, the algebraic multiplicity of 
the zero eigenvalue, $\lambda_0=0$ is given by
\begin{equation}
a_0(t)=n- p_1(t)-1.
\label{eq:a02}
\end{equation}
In this case, 
the corresponding right-eigenvectors $\v(t)$ satisfy
$\alpha(\v(t))=0$, that is, they are orthogonal to $\1$.
\end{description}
\end{thm}
\noindent{\it Proof} \,
In the proof, we will simply write
$p_1=p_1(t)$ and $p_2=p_2(t)$. 
We solve \eqref{eq:ev1b} as follows,
\begin{align*}
\v(t) &= \delta \alpha(\v(t)) 
\{z I -(S^{t+1}+b S^{t+2})  \}^{-1} \1
\nonumber\\
&= \delta \alpha(\v(t)) \sum_{k=0}^{\infty} z^{-(k+1)} 
(S^{t+1}+b S^{t+2})^k \1
\nonumber\\
&= \delta \alpha(\v(t)) 
\sum_{k=0}^{\infty} z^{-(k+1)}
\sum_{q=0}^k \binom{k}{q} b^q S^{(t+1)k+q} \1,
\end{align*}
where we used the expansion formula of an inverse matrix.
By taking inner products with $\1$ on both sides, we have
\[
\bra \v(t), \1 \ket 
=\delta \alpha(\v(t)) \sum_{k=0}^{\infty} z^{-(k+1)} 
\sum_{q=0}^k \binom{k}{q} b^q
\bra S^{(t+1)k+q} \1, \1 \ket.
\]
Using the fact $\bra \v(t), \1 \ket=\alpha(v(t))$ and
the equality
\[
\bra S^{\ell} \1, \1 \ket 
=(n-\ell) 1_{(1 \leq \ell \leq n-1)}
\]
for $S=S_n$, we arrive at the equality,
\[
\alpha(\v(t))= \delta \alpha(\v(t))
\sum_{k=0}^{p_1} z^{-(k+1)}
\sum_{q=0}^{k} \binom{k}{q} b^q
\Big[ n-\{(t+1)k+q\} \Big]
1_{(1 \leq (t+1)k +q \leq n-1)}.
\]
For non-zero eigenvalues, 
we can assume $\alpha(\v(t)) \not=0$. Then we have
\[
\frac{1}{n \delta}
-
\sum_{k=0}^{p_1} z^{-(k+1)}
\sum_{q=0}^{k} \binom{k}{q} b^q
\left\{ 1- \left(\frac{t+1}{n} k + \frac{q}{n} \right)
\right\}
1_{(1 \leq (t+1)k +q \leq n-1)}=0.
\]
The condition $(t+1)k +q \leq n-1$ gives
$q \leq n-1-(t+1)k$.
We see that for $n \in \N$, 
\[
k \leq n-1-(t+1)k, \quad
t \in \{2,3,\dots \}, k \in \N_0:=\{0,1,\dots\} \quad
\iff \quad
k \leq p_2, \quad k \in \N_0
\]
where $p_2$ is defined by \eqref{eq:p2}. 
That is, if $k \geq p_2+1$, then
the summation over $q$ is taken only in the
interval $q \in [0, n-1-(t+1)k]$ instead of $[0, (t+1) k]$.
In other words, in the index space 
$(k, q) \in \N_0 \times \N_0$, 
the region for which we should take summation is
\[
\{(k, q) ; 1 \leq k \leq p_2, 0 \leq q \leq k \} 
\cup
\{(k, q) ; p_2+1 \leq k \leq p_1, 0 \leq q \leq n-1-(t+1)k\}.
\]
This is equivalent with
\[
\{(k, q) ; 1 \leq k \leq p_1, 0 \leq q \leq k \}
\setminus
\{(k, q) ; p_2+1 \leq k \leq p_1, 
n-(t+1)k \leq q \leq k\}.
\]
Hence the above equation is written as
\begin{align}
&\frac{1}{n \delta}
-
\sum_{k=0}^{p_1} z^{-(k+1)}
\sum_{q=0}^{k} \binom{k}{q} b^q
\left\{ 1- \left(\frac{t+1}{n} k + \frac{q}{n} \right)
\right\}
\nonumber\\
& \quad
+
\sum_{k=p_2+1}^{p_1} z^{-(k+1)}
\sum_{q=n-(t+1)k}^{k} \binom{k}{q} b^q
\left\{ 1- \left(\frac{t+1}{n} k + \frac{q}{n} \right)
\right\} =0.
\label{eq:EQ1}
\end{align}
We notice that
\begin{align*}
I_1(k) &:= \sum_{q=0}^{k} \binom{k}{q} b^q
\left\{ 1- \left( \frac{t+1}{n} k +\frac{q}{n} \right) \right\}
= \left( 1 - \frac{t+1}{n} k \right)
\sum_{q=0}^{k} \binom{k}{q} b^q
-\frac{1}{n} \sum_{q=0}^{k} q \binom{k}{q} b^q
\nonumber\\
&= \left( 1 - \frac{t+1}{n} k \right) 
(1+b)^{k}
-\frac{1}{n} bk (1+b)^{k-1}
= (1+b)^{k}
-\frac{1}{n} \frac{(1+b)(t+1) +b}{1+b} k (1+b)^{k},
\end{align*}
and hence
\begin{align*}
J_1 &:= \sum_{k=0}^{p_1} z^{-(k+1)} I_1(k)
\nonumber\\
&=(1+b)^{-1} \left( \frac{z}{1+b} \right)^{-1}
\sum_{k=0}^{p_1}
\left( \frac{z}{1+b} \right)^{-k}
\nonumber\\
& \quad
-\frac{1}{n} \frac{(1+b)(t+1) +b}{1+b} (1+b)^{-1}
\left( \frac{z}{1+b} \right)^{-1}
\sum_{k=0}^{p_1} k
\left( \frac{z}{1+b} \right)^{-k}.
\end{align*}
It is easy to verify that this is written as
\begin{align*}
J_1&=\frac{1}{1+b}
\left(\frac{z}{1+b} \right)^{-(p_1+1)}
\frac{ \displaystyle{1-\left(\frac{z}{1+b} \right)^{p_1+1}} }
{\displaystyle{1-\frac{z}{1+b} } }
\nonumber\\
&\qquad
-\frac{1}{n} \frac{(1+b)(t+1)+b}{(1+b)^2}
\frac{
\displaystyle{
\left(\frac{z}{1+b} \right)^{-(p_1+1)}
}
}{\displaystyle{1-\frac{z}{1+b}}}
\left\{
p_1+1
- 
\frac{ \displaystyle{1-\left(\frac{z}{1+b} \right)^{p_1+1}} }
{\displaystyle{1-\frac{z}{1+b} } }
\right\}.
\end{align*}
We also see that
\begin{align*}
J_2&:= \sum_{k=p_2+1}^{p_1} z^{-(k+1)}
\sum_{q=n-(t+1)k}^{k} b^q \binom{k}{q} 
\left\{ 1- \left(\frac{t+1}{n} k + \frac{q}{n} \right)
\right\} 
\nonumber\\
&=
- \left(\frac{z}{1+b} \right)^{-(p_1+1)}
\frac{1}{(1+b)^{p_1+1}}
\sum_{k=p_2+1}^{p_1} z^{p_1-k}
\sum_{q=n-(t+1)k+1}^{k} b^q \binom{k}{q} 
\frac{1}{n} \{q-(n-(t+1)k)\}
\nonumber\\
&=
- \left(\frac{z}{1+b} \right)^{-(p_1+1)}
\frac{1}{(1+b)^{p_1+1}}
\sum_{\ell=0}^{p_1-p_2-1} z^{\ell}
\nonumber\\
& \qquad \times
\sum_{q=n-(t+1)(p_1-\ell)+1}^{p_1-\ell} b^q
\binom{p_1-\ell}{q}
\frac{1}{n} [ q-\{n-(t+1)(p_1-\ell)\}], 
\end{align*}
where we have noticed that
the term with $q=n-(t+1)k$ vanishes 
in the first line, and then
the summation over $k$ is replaced by
the summation over $\ell:=p_1 -k$ at the last equality.
$J_2$ should be zero if $p_1=p_2$.
For $0 \leq \ell \leq p_1-p_2-1$,
$n-(t+1) p_1+1 \leq n-(t+1)(p_1-\ell)+1
\leq q \leq p_1-\ell \leq p_1$
in the second summation in the last line. 
Hence, if $n-(t+1) p_1+1>p_1 \iff
p_1 < (n+1)/(t+1+1)$, then $J_2$ should be zero.
Equation \eqref{eq:EQ1} is thus given by
\[
\frac{1}{n \delta}
- J_1+1_{(p_1 \geq p_2+1, \, p_1 \geq (n+1)/(t+2))} J_2=0.
\]
Multiply the factor
$(1+b) \times (z/(1+b))^{p_1+1}$.
Then \eqref{eq:eigenvaluesG1} is obtained.
It is easy to show that this equation is
also written as \eqref{eq:eigenvalues}.
\qed

\vskip 0.3cm
\begin{rem}
\label{rem:remark1}
The last term in the left-hand side of
\eqref{eq:eigenvalues} will vanish or will be reduced into
only one term if $n \geq \lceil n-1 \rceil$, where
$\lceil x \rceil$ denotes the smallest integer greater 
than or equal to
$x \in \R$ (the ceiling function of $x$). 
The following is 
verified \cite[Lemma 3.6]{MKS24}.
Let
$I_{n-1}:=[ \lceil \sqrt{n-1} \, \rceil, n-1 ] \cap \N$ 
and
$T_{n-1}:=\{ \lfloor (n-1)/k \rfloor ; k=1,2, \dots, n-1\}$.
Then
\[
p_1(t)-p_2(t)=\begin{cases}
1, \quad \mbox{if $t+1 \in I_{n-1} \cap T_{n-1}$},
\cr
0, \quad \mbox{if $t+1 \in I_{n-1} \setminus T_{n-1}$}.
\end{cases}
\]
Moreover, when $b=0$, \eqref{eq:eigenvalues} 
is reduced to much simpler form as
\begin{equation}
z^{p_1(t)+1}
-n \delta \sum_{k=0}^{p_1(t)} \left\{
1-(p_1(t)-k) \frac{t+1}{n} \right\} z^k =0.
\label{eq:eigenvalues0}
\end{equation}
\end{rem}
The following are corollaries
of Theorem \ref{thm:ev1} 
(see \cite{MKS24} for more detail).

\begin{prop}
\label{thm:ev2}
There is an outlier eigenvalue $\lambda_1(t)$, whose modulus 
goes to $\infty$ as $n \delta \to \infty$.
For $n \delta >1$,
we have the expression 
\begin{align}
\lambda_1(t) &=n \delta+ 1 +b-
(1+b) \sum_{k=0}^{p_1(t)-1} C_k 
\left( \frac{t+1}{n}+\frac{b}{(1+b) n}
\right)^{k+1} 
\left( \frac{1+b}{n \delta} \right)^k
+\rO((n \delta)^{-p_1(t)})
\nonumber\\
& \qquad \qquad 
+1_{(p_1(t) \geq p_2(t)+1)} \rO((n \delta)^{-p_2(t)}),
\label{eq:large_evG}
\end{align}
where $C_k$ is the Catalan number;
$\displaystyle{C_k := \frac{1}{k+1} \binom{2k}{k}
=\frac{(2k)!}{(k+1)! k!}}$, 
$k \in \N_0$.
\end{prop}
\begin{prop}
\label{thm:ev3}
Fix $\delta$ and $b$. Consider $t$ satisfying
$\delta > 4\{ (1+b) (t+1) + b \}/n^2$ 
and $p_1(t) =p_2(t)$. 
Then as $n \to \infty$, $p_1(t)$ non-zero eigenvalues 
except the outlier eigenvalue $\lambda_1(t)$ 
tend to form a configuration such that
one point at $z=1+b$ is eliminated from the 
equidistance $p_1(t)+1$ points
on a circle centered the origin with radius $1+b$; 
\[
(1+b) e^{2 \pi i \ell/(p_1(t)+1)}, \quad \ell=0,1, \dots, p_1(t).
\]
\end{prop}

\noindent
For \textit{outliers in spectra} discussed 
in random matrix theory,
see \cite{For23,Tao13}.

For the non-zero eigenvalues $\lambda_{j}(t)$,
$j=1,2, \dots, p_1(t)+1$, 
the right- and left-eigenvectors are denoted by
$\v_j(t)$ and $\w_j(t)$, respectively,
$t=0, 1, \dots, T$.
They are determined to satisfy the 
\textit{bi-orthogonality relation},
\begin{equation}
\bra \w_j(t), \v_k(t) \ket=\delta_{j, k},
\quad j, k=1,2, \dots, p_1(t)+1,
\quad t=0, 1, \dots, T.
\label{eq:eigenvectors}
\end{equation}

\SSC
{Pseudospectrum Dynamics}
\label{sec:PS}
\subsection{Generalized eigenspaces associated with
$\lambda_0=0$}
\label{sec:generalES}
The eigenvalue and right-eigenvector equations
\eqref{eq:ev1} with \eqref{eq:model}, 
\[
(S^{t+1}+ bS^{t+2}+ \delta J) \v(t)=\lambda(t) \v(t), 
\]
is written as
\[
(S^{t+1}+ b S^{t+2}) \v(t) =\lambda(t) \v(t)
- \delta \alpha(\v(t)) \1,
\]
where $\alpha(\v(t))$ was defined by \eqref{eq:alpha}
and we have used the equality
\eqref{eq:Jv=a1}.
We consider the right-eigenvectors $\v_0(t)$ associated 
with the zero-eigenvalue $\lambda(t) \equiv \lambda_0=0$;
\[
(S^{t+1}+ b S^{t+2}) \v_0(t) =- \delta \alpha(\v_0(t)) \1,
\quad t=1,2, \dots, T.
\]
Let $\e_j :=(\underbrace{0, \cdots, 0}_{j-1}, 1, 
\underbrace{0, \cdots, 0}_{n-j})^{\sT} \in \C^n$
be the $j$-th standard vector, $j=1,2, \dots, n$, 
and $V_k:={\rm span}\{\e_1, \e_2, \dots, \e_k\}$,
$k=1,2, \dots, n$.
Since $S^{t+1}$ and $S^{t+2}$ are upward shift operators
for vectors by $t+1$ and $t+2$, respectively,
$(S^{t+1}+ b S^{t+2})\u \in V_{n-(t+1)}$
for any $\u \in \C^n$.
On the other hand, $\1$ is the all-ones vector in $V_n$,
and hence, $\1 \notin V_{n-(t+1)}$.
Therefore, $\alpha(\v_0(t))=0$ is concluded.
The right-eigenvectors 
$\v_0(t)=(v_{01}(t), \dots, v_{0n}(t))^{\sT}$ for $\lambda_0$ 
are hence the solutions of the equations
\begin{align}
&(S^{t+1}+b S^{t+2}) \v_0(t)=0,
\label{eq:v01a}
\\
& \sum_{j=1}^n v_{0j}(t)=0,
\quad t=1, 2, \dots, T.
\label{eq:v01b}
\end{align}

\begin{prop}
\label{thm:g0}
The geometric multiplicity of $\lambda_0$ is given by
\begin{equation}
g_0(t)=t.
\label{eq:g02}
\end{equation}
\end{prop}
\noindent \textit{Proof} \,
It is easy to verify that all the vectors 
satisfying \eqref{eq:v01a} are in $V_{t+1}$.
Since the zero-sum condition \eqref{eq:v01b} is
imposed for $v_{0j} \in \C, j=1, \dots, t+1$,
the dimension of complex right-eigenvectors $\v_0$ is
equal to $t$. The proof is complete. \qed
\vskip 0.3cm
At each time $t=1, 2, \dots, T$, 
we write the $t$ linearly independent 
solutions of \eqref{eq:v01a} with \eqref{eq:v01b}
as $\v_{0}^{(\ell, 1)}(t)$, $\ell=1, 2, \dots, t$.
We choose the following solutions, 
\begin{equation}
\v_0^{(\ell, 1)}(t)
=\e_1-\e_{\ell+1}
=\Big(1, \underbrace{0, \cdots, 0}_{\ell-1}, -1,
\underbrace{0, \cdots, 0}_{n-(\ell+1)}
\Big)^{\sT} 
\in V_{\ell+1} \subset V_{t+1}.
\label{eq:v0_ell1}
\end{equation}
Assume that $t=1,2, \dots, T-1$.
For each right-eigenvector $\v_0^{(\ell, 1)}(t)$,
$\ell=1,2, \dots, t$, the 
\textit{generalized right-eigenvectors},
$\v_0^{(\ell, q)}(t)$, 
are given by the solutions of
the \textit{Jordan chain},
\begin{align}
& (S^{t+1}+b S^{t+2} + \delta J) 
\v_0^{(\ell, q)}(t)=\v_0^{(\ell, q-1)}(t)
\nonumber\\
\iff \quad
&(S^{t+1}+b S^{t+2}) \v_0^{(\ell,q)}(t)
=\v_0^{(\ell, q-1)}(t) - \delta \alpha(\v_0^{(\ell, q)}) \1,
\quad q=2,3, \dots.
\label{eq:general_evec1}
\end{align}
Since $(S^{t+1}+bS^{t+2}) \v_0^{(\ell, q)}(t) \in V_{n-(t+1)}$
in general, $\v_0^{(\ell, 1)}(t) \in V_{t+1}$ 
implies $\alpha(\v_0^{(\ell, 2)})$ multiplied by $\1$
should be zero.
By reduction with respect to $q$, 
we show that the Jordan chain
\eqref{eq:general_evec1} is equivalent with
\begin{align}
&(S^{t+1}+b S^{t+2}) 
\v_0^{(\ell, q)}(t)=\v_0^{(\ell, q-1)}(t), 
\label{eq:general_evec2a}
\\
&\sum_{j=1}^n v_{0j}^{(\ell, q)}(t)=0, 
\quad q=2,3, \dots.
\label{eq:general_evec2b}
\end{align}
The dimension of the generalized 
$\ell$-th sub-eigenspace denoted by
$d_0^{(\ell)}(t)$ is the highest value of $q$
satisfying \eqref{eq:general_evec2a} 
and \eqref{eq:general_evec2b} with
non-zero $\v_0^{(\ell, q)}(t)$.
We put
\[
k_0(t):= \max_{\ell: 1 \leq \ell \leq t}
d_0^{(\ell)}(t),
\quad
t=1,2, \dots, T,
\]
and remember that $\xi(t,n)$ is defined by 
\eqref{eq:xi}.

\begin{prop}
\label{thm:k0}
Assume that $t=1,2, \dots, T-1$. 
\begin{description}
\item{\rm (i)} \,
If $\xi(t, n) \in \{0,1\}$, then
\begin{equation}
d_0^{(\ell)}(t)=\left\lfloor \frac{n}{t+1} \right\rfloor,
\quad \ell=1,2, \dots, t.
\label{eq:d0A}
\end{equation}
Hence
\[
k_0(t)= \left\lfloor \frac{n}{t+1} \right\rfloor.
\]
\item{\rm (ii)} \,
If $\xi(t,n) \in \{2,3,\dots,t\}$, then
\begin{equation}
d_0^{(\ell)}(t)= 
\begin{cases}
\displaystyle{\left\lfloor \frac{n}{t+1} \right\rfloor+1},
\quad
& 1 \leq \ell \leq \xi(t,n)-1,
\cr
\cr
\displaystyle{\left\lfloor \frac{n}{t+1} \right\rfloor},
\quad
& \xi(t,n) \leq \ell \leq t.
\end{cases}
\label{eq:d0B}
\end{equation}
Hence
\[
k_0(t)= \left\lfloor \frac{n}{t+1} \right\rfloor+1.
\]
\end{description}
\end{prop}
\vskip 0.3cm
\noindent{\it Proof} \,
First consider \eqref{eq:general_evec2a} for $q=2$.
Since $S^{t+1}+bS^{t+2}=(I+b S) S^{t+1}$,
and $I+bS$ is invertible, 
we have
\begin{equation}
S^{t+1} \v_0^{(\ell, 2)}(t)=(I+bS)^{-1} \v_0^{(\ell, 1)}(t).
\label{eq:chain1}
\end{equation}
For $\u=(u_1, \dots, u_n)^{\sT} \in V_n$,
we define
\begin{equation}
S^{-(t+1)} \u=
( \underbrace{0, \cdots, 0}_{t+1}, 
u_1, u_2, \dots, u_{n-(t+1)})^{\sT}. 
\label{eq:S_inverse}
\end{equation}
Then \eqref{eq:chain1} is solved by
\begin{equation}
\v_0^{(\ell,2)}(t)=\u_1+\u_2,
\label{eq:chain1b}
\end{equation}
where 
$\u_1$ is an arbitrary vector 
in $V_{t+1}$ and
$\u_2:=S^{-(t+1)}(I+bS)^{-1} \v_0^{(\ell,1)}(t)$.
By the fact $\v_0^{(\ell, 1)}(t) \in V_{\ell+1}$,
we see that
\[
(I+bS)^{-1} \v_0^{(\ell, 1)}(t)
=\sum_{k=0}^{\ell} (-b)^{k} S^{k} \v_0^{(\ell, 1)}(t)
\in V_{\ell+1}.
\]
Hence, 
$\v_0^{(\ell, 2)}(t) \in V_{(t+1)+(\ell+1)}$ is
written in the form
\[
\v_0^{(\ell, 2)}(t)=
(u_{11}, \dots, u_{1 \, t+1}, u_{21}, \dots, u_{2 \,\ell+1},
\underbrace{0, \cdots, 0}_{n-(t+1)-(\ell+1)}
)^{\sT},
\]
provided $\ell+1 \leq n-(t+1)$. 
Finally, 
$\u_1=(u_{11}, \dots, u_{1 \, t+1},
\underbrace{0, \cdots, 0}_{n-(t+1)})^{\sT} \in V_{t+1}$
is chosen to satisfy 
$\alpha(\u_1)=-\alpha(\u_2)$, that is, to satisfy
the zero-sum condition 
\eqref{eq:general_evec2b} for $q=2$.
One possibility is that 
\begin{equation}
\u_1=-\alpha(\u_2) \e_1 \in V_1.
\label{eq:u1}
\end{equation}
By the same procedure, 
we can determine 
$\v_0^{(\ell, q)}(t) \in V_{(q-1)(t+1)+(\ell+1)}$
from $\v_0^{(\ell, q-1)}(t) \in V_{(q-2)(t+1)+(\ell+1)}$ for 
$q=2, 3, \dots, d_0^{(\ell)}(t)$, where
$d_0^{(\ell)}(t)$ is given by 
\eqref{eq:d0A} if $\xi(t,n) \in \{0,1\}$, 
and by \eqref{eq:d0B}
if $\xi(t,n) \in \{2,3,\dots,t\}$, respectively.
The proof is complete.
\qed
\vskip 0.3cm

Next we consider the 
left-eigenvectors for $\lambda_0$.
Following the similar argument 
for the right-eigenvectors, we can show that 
they are given by the solutions of the following, 
\begin{align}
&\w_0(t)^{\dagger}(S^{t+1}+b S^{t+2})=0,
\label{eq:w01a}
\\
& \sum_{j=1}^n w_{0j}(t)=0,
\quad t=1, 2, \dots, T.
\label{eq:w01b}
\end{align}
The general form of the 
left-eigenvector $\w_0(t)$ 
is hence given by
\begin{equation}
\w_0(t)=\Big(
\underbrace{0, \cdots, 0}_{n-(t+1)},
w_{0, n-t}(t), \dots, w_{0, n}(t) \Big)^{\sT}
\label{eq:w02}
\end{equation}
with the condition \eqref{eq:w01b}. 
For a given set of generalized right-eigenvectors
$\{\v_0^{(\ell, q)}(t) : 
\ell=1,2, \dots, t, q=1, \dots, d_0^{(\ell)}(t) \}$, 
first we determine the left-eigenvectors
\eqref{eq:w02} with \eqref{eq:w01b},
which we write as $\w_0^{(\ell, d_0^{(\ell)}(t))}(t)$,
$\ell=1,2, \dots, t$ by convention, 
so that they satisfy the \textit{bi-orthonormality relation}, 
\begin{equation}
\bra \w_0^{(\ell, d_0^{(\ell)}(t))}(t), 
\v_0^{(r, q)}(t) \ket
=\delta_{\ell, r} \delta_{d_0^{(\ell)}(t), q}, \quad
\ell, r = 1,2, \dots, t,
\quad q=1,2, \dots, d_0^{(\ell)}(t). 
\label{eq:orthogonal1}
\end{equation}
As shown above for the generalized right-eigenvectors,
we can also prove the following for $t=1,2, \dots, T-1$. 
For each left-eigenvector $\w_0^{(\ell, d_0^{(\ell)}(t))}(t)$,
$\ell=1, 2, \dots, t$, 
the generalized left-eigenvectors,
$\w_0^{(\ell, d_0^{(\ell)}(t)-q)}(t)$, $q=1, 2, \dots, 
d_0^{(\ell)}(t)-1$, 
are given by the solutions of the following Jordan chain, 
\begin{align}
&{\w_0^{(\ell, d_0^{(\ell)}(t)-q)}}(t)^{\dagger}
(S^{t+1}+b S^{t+2}) 
={\w_0^{(\ell, d_0^{(\ell)}(t)-(q-1))}}(t)^{\dagger},
\nonumber\\
&\sum_{j=1}^n w_{0 j}^{(\ell, d_0^{(\ell)}(t)-q)}(t) =0,
\quad q=1,2, \dots, d_0^{(\ell)}(t)-1.
\label{eq:model2_g_e_vec1_left}
\end{align}
They are uniquely determined by the 
bi-orthonormality condition,
\begin{equation}
\bra \w_0^{(\ell, p)}(t), \v_0^{(r, q)}(t) \ket =
\delta_{\ell r} \delta_{p, q}, \quad
\ell, r=1,2, \dots, t, \quad
p, q=1, \dots, d_0^{(\ell)}(t).
\label{eq:orthogonal2}
\end{equation}

\begin{rem}
\label{thm:remark3_1}
Although the right-eigenvalues 
$\v_0^{(\ell, 1)}(t)$, $\ell=1,2, \dots, t$
are simple as shown by \eqref{eq:v0_ell1},
explicit expressions
for the generalized right-eigenvectors  
$\v_0^{(\ell, q)}(t)$, $q=2, 3, \dots, d_0^{(\ell)}(t)$
become more complicated. 
For an example, here we show the results
for $n=6$ at time $t=2$.
There are $t=2$ right-eigenvectors in the simple form,
$\v_0^{(1,1)}=(1, -1, 0, 0, 0, 0)^{\sT}$ and
$\v_0^{(2,1)}=(1, 0, -1, 0, 0, 0)^{\sT}$.
Since $d_0^{(1)}(2)=d_0^{(2)}(2)=\lfloor 6/(2+1) \rfloor=2$, 
we have one generalized right-eigenvector
for each as
\[
\v_0^{(1,2)}(2)=(-b, 0, 0, 1+b, -1, 0)^{\sT}, \quad
\v_0^{(2,2)}(2)=(-b+b^2, 0, 0, 1-b^2, b, -1)^{\sT},
\]
respectively.
Here we have followed the choice \eqref{eq:u1}.
The generalized left-eigenvectors are 
then determined uniquely by solving 
the Jordan chain \eqref{eq:model2_g_e_vec1_left}
with the orthonormality condition
\eqref{eq:orthogonal2} as
\begin{align*}
\w_0^{(1,1)}(2) &=
\left(\frac{1+b}{3+2b}, -\frac{2+b}{3+2b}, \frac{1+b}{3+2b},
\frac{b(1+2b)}{(3+2b)^2}, - \frac{2b(1+b)}{(3+2b)^2},
-\frac{2b(1+b)}{(3+2b)^2} \right)^{\dagger},
\nonumber\\
\w_0^{(1,2)}(2) &=
\left(0, 0, 0, \frac{1+b}{3+2b}, -\frac{2-b^2}{3+2b},
\frac{1-b-b^2}{3+2b} \right)^{\dagger},
\nonumber\\
\w_0^{(2,1)}(2) &=
\left( \frac{1}{3+2b}, \frac{1}{3+2b}, - \frac{2(1+b)}{3+2b},
\frac{4b}{(3+2b)^2}, \frac{b(1+2b)}{(3+2b)^2}, 
\frac{b(1+2b)}{(3+2b)^2} \right)^{\dagger},
\nonumber\\
\w_0^{(2,2)}(2) &=
\left( 0, 0, 0, \frac{1}{3+2b}, \frac{1+b}{3+2b}, 
-\frac{2+b}{3+2b} \right)^{\dagger}.
\end{align*}
\end{rem}

\begin{rem}
\label{thm:remark3_2}
At each time $t=1,2, \dots, T-1$, 
the first $(q=2)$ generalized right-eigenvector
is given by
\begin{align}
&\v_0^{(\ell,2)}(t)
\nonumber\\
&
=\Big(
\frac{b\{-1+(-b)^{\ell}\}}{1+b}, 
\underbrace{0, \cdots, 0}_{t},
1-(-b)^{\ell}, -(-b)^{\ell-1}, \cdots, -(-b), -1,
\underbrace{0, \cdots, 0}_{n-t-\ell-2}
 \Big)^{\sT},
\label{eq:v0_ell2}
\end{align}
provided $n \geq (t+1)+(\ell+1)$,
where the choice \eqref{eq:u1} was taken.
The proof of \eqref{eq:v0_ell2} is given in
a general setting in Appendix \ref{sec:general}.
When $b=0$, this is reduced 
to be the inverse shifts of $\v_0^{(\ell, 1)}(t)$
given by \eqref{eq:v0_ell1}, 
\begin{align*}
\v_0^{(\ell, 2)}(t)
&=S^{\sT} \v_0^{(\ell, 1)}(t)
\nonumber\\
&= ( \underbrace{0, \cdots, 0}_{t+1},
1, \underbrace{0, \cdots, 0}_{\ell-1}, -1,
\underbrace{0, \cdots, 0}_{n-t-\ell-2}).
\end{align*}
Explicit expressions for the generalized 
right- and left-eigenvectors
are given for the case $b=0$ in Appendix \ref{sec:b0}.
\end{rem}

\subsection{Jordan decomposition of resolvent
and generalized condition numbers}
\label{sec:Jordan}
For each time $t = 1, 2,\dots, T$ 
and $\ell = 1,2, \dots, t$, 
define the $n \times d_0^{(\ell)}(t)$ rectangular matrices
\begin{align*}
V_{0, \ell}(t) &:=
\Big(
\v_0^{(\ell, 1)}(t) \, \v_0^{(\ell, 2)}(t) \, \cdots \,
\v_0^{(\ell, d_0^{(\ell)}(t))}(t) \Big), 
\nonumber\\
W_{0, \ell}(t) &:=
\Big(
\w_0^{(\ell, 1)}(t) \, \w_0^{(\ell, 2)}(t) \, \cdots \,
\w_0^{(\ell, d_0^{(\ell)}(t))}(t) \Big).
\end{align*}
Then, by the construction of the generalized
eigenspaces given in Section \ref{sec:generalES}, we have
\begin{equation}
S^{(b)}(t, \delta J)
= \sum_{\ell=1}^{t} V_{0, \ell}(t)
S_{d_0^{(\ell)}(t)} W_{0, \ell}(t)^{\dagger}
+ \sum_{j=1}^{p_1(t)+1} \lambda_j(t) P_j(t),
\label{eq:general_VW}
\end{equation}
at each time $t=1,2, \dots, T$, 
where we have defined the projection operators as
\begin{equation}
P_j(t) := \v_j(t) {\w_j(t)}^{\dagger} \in \C^{n \times n}, 
\quad j=1,2, \dots, p_1(t)+1.
\label{eq:Pk}
\end{equation}

\begin{rem}
\label{thm:remark3_3}
For each time $t=1,2, \dots, T$, define
the $n \times n$ matrices
\begin{align*}
V(t) &:= \Big( V_{0, 1}(t) \, V_{0, 2}(t) \,
\cdots V_{0, t}(t) \,
\v_1(t) \, \v_2(t) \cdots \, \v_{p_1(t)+1} \Big),
\nonumber\\
W(t) &:= \Big( W_{0, 1}(t) \, W_{0, 2}(t) \,
\cdots W_{0, t}(t) \,
\w_1(t) \, \w_2(t) \cdots \, \w_{p_1(t)+1} \Big).
\end{align*}
Then 
\begin{equation}
W(t)=V(t)^{-1}, 
\label{eq:V-1}
\end{equation}
and 
\[
W(t) S^{(b)}(t, \delta J) V(t)
=\J(t) \oplus
{\rm diag} \Big (\lambda_1(t), \lambda_2(t), \cdots, 
\lambda_{p_1(t)+1}(t) \Big), 
\quad
t=1, 2, \dots, T,
\]
where
\[
\J(t)=\bigoplus_{\ell=1}^t S_{d_0^{(\ell)}(t)}.
\]
In order to calculate $W(t)$ by \eqref{eq:V-1}, 
we need to know the exact values of non-zero eigenvalues
$\lambda_j(t)$ and
their eigenvectors $\v_{j}(t)$, $j=1, 2, \dots, p_1(t)+1$.
The method using bi-orthonormality relations
explained in Section \ref{sec:generalES}
gives the parts of $W(t)$ associated with $\lambda_0$;
$W_{0, \ell}(t)$, $\ell=1, 2, \dots, t$, 
without these information.
\end{rem}

Let $I$ be the $n \times n$ identity matrix. 
The \textit{resolvents} 
of $(S^{(b)}(t, \delta J))_{1 \leq t \leq T}$ 
are then expanded as
\begin{align}
\Big(z I-S^{(b)}(t, \delta J) \Big)^{-1}
&=  \sum_{\ell=1}^{t}
V_{0, \ell}(t) \Big(z I-S_{d_0^{(\ell)}(t)} \Big)^{-1} 
W_{0, \ell}(t)^{\dagger}
+\sum_{j=1}^{p_1(t)+1} \frac{1}{z-\lambda_j(t)} P_j(t)
\nonumber\\
&=
\sum_{\ell=1}^{t}
V_{0, \ell}(t)
\left( \frac{1}{z} I + \frac{1}{z^2} S_{d_0^{(\ell)}(t)}
+ \cdots + \frac{1}{z^{d_0^{(\ell)}(t)}} 
S_{d_0^{(\ell)}(t)}^{d_0^{(\ell)}(t)-1}
\right) W_{0, \ell}(t)^{\dagger}
\nonumber\\
& \quad
+\sum_{j=1}^{p_1(t)+1} \frac{1}{z-\lambda_j(t)} P_j(t),
\quad t=1, 2,  \dots, T, 
\label{eq:resolvent1}
\end{align}
This formula is a special case of Eq.~(5.23) 
on page 40 in \cite{Kato80}.
For each $\ell \in \{1,2, \dots, t\}$,
$S_{d_0^{(\ell)}(t)}^{d_0^{(\ell)}(t)-q} 
\in \C^{d_0^{(\ell)}(t) \times d_0^{(\ell)}(t)}$
with $q \in \{1,2, \dots, d_0^{(\ell)}(t)\}$
is a matrix whose $(j,k)$-entries are 1 if
$k-j=d_0^{(\ell)}(t)-q$, and are zero otherwise.
Hence \eqref{eq:resolvent1} is written as
\begin{align}
&\Big(z I-S^{(b)}(t, \delta J) \Big)^{-1}
\nonumber\\
&\qquad =
\sum_{\ell=1}^{t} \sum_{q=1}^{d_0^{(\ell)}(t)}
\frac{1}{z^{d_0^{(\ell)}(t)-q+1}}
\Big( \v_0^{(\ell, 1)}(t) \, \v_0^{(\ell, 2)}(t) \,
\cdots \, \v_0^{(\ell, q)}(t) \Big)
\left(
\begin{matrix}
\w_0^{(\ell, d_0^{(\ell)}(t)-q+1)}(t)^{\dagger} \cr
\w_0^{(\ell, d_0^{(\ell)}(t)-q+2)}(t)^{\dagger} \cr
\cdots \cr
\w_0^{(\ell, d_0^{(\ell)}(t))}(t)^{\dagger}
\end{matrix}
\right)
\nonumber\\
& \qquad
+\sum_{j=1}^{p_1(t)+1} \frac{1}{z-\lambda_j(t)} P_j(t),
\quad t=1, 2,  \dots, T. 
\label{eq:resolvent1b}
\end{align}

For a matrix $M \in \C^{n \times n}$, the 2-norm of $M$ is defined by
\[
\| M \| :=
\max_{\x \in \C^n} \frac{\| M \x\|}{\|\x\|}
=\max_{\x \in \C^n: \|\x\|=1} \|M \x \|, 
\]
where $\| \x \|$ denotes the 2-norm of vector $\x$.
Therefore, for $t=1, 2, \dots, T$, 
as $z \to \lambda_j(t)$, $j=1,2, \dots, p_1(t)+1$, 
\[
\|(z I-S^{(b)}(t, \delta J))^{-1} \| 
=
|z-\lambda_j(t)|^{-1} \| P_j(t) \|
+\rO(1)
\]
and as $z \to 0$, 
\[
\|(z I-S^{(b)}(t, \delta J))^{-1} \| 
\leq
|z|^{-k_0(t)}
\sum_{\ell: d_0^{(\ell)}(t)=k_0(t)}
\|\v_0^{(\ell,1)}(t) \w_0^{(\ell, d_0^{(\ell)}(t))}(t)^{\dagger} \|
+\rO(|z|^{-k_0(t)+1}).
\]
We see that
\[
\|P_j(t) \| 
=\|\v_j(t) \w_j(t)^{\dagger}\|
=\| \v_j(t) \| \| \w_j(t) \| =: \kappa_j(t),
\]
where the value $\kappa_j(t)$ is 
know as the
\textit{eigenvalue condition number}
of $\lambda_j(t)$,
$j=1, 2, \cdots, T$ \cite{TE05}. 
We also see that 
\begin{equation}
\|\v_0^{(\ell, 1)}(t) 
\w_0^{(\ell, d_0^{(\ell)}(t))}(t)^{\dagger} \|
=\|\v_0^{(\ell, 1)}(t)\|
\|\w_0^{(\ell, d_0^{(\ell)}(t))}(t)\|
=: \kappa_0^{(\ell)}(t), 
\label{eq:generalCN}
\end{equation}
which we may call 
the \textit{generalized condition number}
of $\lambda_0$ associated with
its $\ell$-th Jordan block such that
$d_0^{(\ell)}(t)=k_0(t)$.
We put
\[
\kappa_0(t) := \max_{\ell: d_0^{(\ell)}(t)=k_0(t)}
\kappa_0^{(\ell)}(t).
\]
By Proposition \ref{thm:k0},
if $\xi(t,n) \in \{0,1\}$,
$k_0(t)=\lfloor n/(t+1) \rfloor$,  
while if $\xi(t,n) \in \{2,3,\dots,t\}$,
$k_0(t)=\lfloor n/(t+1) \rfloor+1$.
Hence
\[
k_0(t) \leq \frac{n+t+1}{t+1}.
\]
Moreover, Proposition \ref{thm:k0} gives that,
if $\xi(t,n) \in \{0,1\}$, 
\[
\sharp \{\ell: d_0^{(\ell)}(t)=k_0(t) \} 
=\sharp \{\ell: 1 \leq \ell \leq t \} =t, 
\]
and 
if $\xi(t,n) \in \{2,3,\dots,t\}$, 
\[
\sharp \{\ell: d_0^{(\ell)}(t)=k_0(t) \} 
=\sharp \{\ell: 1 \leq \ell 
\leq \xi(t,n) -1 \} 
\leq t.
\]
Then we have obtained the following estimates. 
See \cite[Section 52]{TE05}.
\begin{prop}
\label{thm:resolvent1}
For each time $t =1,2, \dots, T$, 
there exist constants $r_j(t) >0$ 
and $C_j(t) >0$,
$j=0, 1, \dots, p_1(t)+1$ such that 
\begin{equation}
\|(z I-S^{(b)}(t, \delta J))^{-1} \|
\leq C_0(t) |z|^{-(n+t+1)/(t+1)}
\label{eq:resolvent3}
\end{equation}
for all $z$ satisfying $|z| \leq r_0(t)$,
and 
\begin{equation}
\|(z I-S^{(b)}(t, \delta J))^{-1} \|
\leq C_j(t) |z-\lambda_j(t)|^{-1}
\label{eq:resolvent4}
\end{equation}
for all $z$ satisfying $|z-\lambda_j(t)| \leq r_j(t)$,
$j=1,2, \dots, p_1(t)+1$.
The infimum of possible value
for $C_0(t)$ is $t \kappa_0(t)$,
and those for $C_j(t)$ are 
$\kappa_j(t)$, $j=1,2, \dots, p_1(t)+1$. 
\end{prop}

\subsection{Time-dependent pseudospectra}
\label{sec:Pseudo}

The singularities of resolvent are exactly eigenvalues 
of the matrix. In \eqref{eq:resolvent1}, the non-zero
eigenvalues $\lambda_j(t), j=1,2, \dots, p_1+1$ give
simple poles of the resolvent.
The singularity at the zero-eigenvalue $\lambda_0$
is realized as the superpositions of poles
with orders from 1 to $d_0^{(\ell)}(t)$ for
$\ell=1, 2, \dots, t$ due to the Jordan-block
structure of the generalized eigenspace associated with
$\lambda_0$. 
This fact implies that the 2-norm of resolvent
takes large values even at the places in $\C$
far from $\lambda_0$.
The set of the points on $\C$ which are not
necessarily eigenvalues exactly but at which the 2-norm
of resolvent becomes large is called
the \textit{pseudospectrum}.

\begin{df}[\cite{RT92,TE05}]
\label{thm:PS}
Let $M \in \C^{n \times n}$ and $\varepsilon >0$
be arbitrary. The $\varepsilon$-pseudospectrum
$\sigma_{\varepsilon}(M)$ of $M$ is the set of
$z \in \C$ such that
\begin{equation}
\| (z I - M)^{-1} \| > \varepsilon^{-1}.
\label{eq:PS1}
\end{equation}
\end{df}
\vskip 0.3cm

Hence, the comparison of 
Proposition \ref{thm:resolvent1} with
Definition \ref{thm:PS} gives the following.
For $\zeta \in \C$ and $r >0$, let
$\D(\zeta, r) := \{z \in \C: |z-\zeta| < r\}$.

\begin{thm}
\label{thm:upperPS}
Let $0 < \varepsilon < 1$. 
For each time $t =1,2, \dots, T$, 
there exist constants $C_j(t) >0$,
$j=0, 1, \dots, p_1(t)+1$ such that 
\begin{equation}
\sigma_{\varepsilon}(S^{(b)}(t, \delta J))
\subseteq
\D \Big(0, 
(\varepsilon C_0(t))^{(t+1)/(n+t+1)} \Big)
\bigcup_{j=1}^{p_1(t)+1} 
\D\Big(\lambda_j(t), \varepsilon C_j(t) \Big).
\label{eq:upperPS}
\end{equation}
The infimum of possible value
for $C_0(t)$ is $t \kappa_0(t)$,
and those for $C_j(t)$ are $\kappa_j(t)$,
$j=1,2,\dots$, $p_1(t)+1$. 
\end{thm}

\begin{rem}
\label{thm:remarks_PS}
\begin{description}
\item{\rm (i)} \,
With fixed $n$ and $t$, as $\varepsilon \to 0$,
the $\varepsilon$-pseudospectrum
including $\lambda_0$ shrinks to the origin.
For large $n$, the exponent $(t+1)/(n+t+1)$ becomes 
small positive value.
Hence, the pseudospectrum including $\lambda_0$
should become a singleton $\{\lambda_0\}$ in the limit 
$\varepsilon \to 0$, as a matter of course, but 
the size reduction of the pseudospectrum
will be very slow for large $n$.

\item{\rm (ii)} \,
With fixed $n$, as time $t$ increases
from 1 to $T$,
$(t+1)/(n+t+1)$ increases monotonically 
from $2/(n+2)$ to 
$(n-1)/(2n-1) \simeq 1/2$.
This suggests that,
if
$\varepsilon < \{\max_{t:1 \leq t \leq T} t \kappa_0(t)\}^{-1}$,
the $\varepsilon$-pseudospectrum
including $\lambda_0$ shrinks
as time $t$ is increasing.
Notice that as given by \eqref{eq:tkappa0_b0},
when $b=0$, 
\[
\left\{\max_{t: 1 \leq t \leq T} t \kappa_0(t) \right\}^{-1}
=\{t \kappa_0(t) |_{t=T}\}^{-1}
=\{2 (n-2)(n-1) \}^{-1/2}. 
\]
This behaves as $n^{-1}/\sqrt{2}$ as $n \to \infty$. 
The size reduction of the $\varepsilon$-pseudospectrum
including $\lambda_0$ in time shows 
the relaxation of defectivity of $\lambda_0$.
For $b \not=0$, explicit calculation of the
generalized condition numbers to evaluate
$t \kappa_0(t)$ is a future problem
(see the item (2) in Sectin \ref{sec:future}). 

\item{\rm (iii)} \,
On the other hand, with fixed time $t$, as 
the size of matrix $n$ increases,
$(t+1)/(n+t+1) \searrow 0$.
This implies that for a given 
$\varepsilon \in (0, 1)$, 
the $\varepsilon$-pseudospectrum
including $\lambda_0$ 
expands as $n \to \infty$. 
\end{description}
\end{rem}

\SSC
{Numerical Analysis}
\label{sec:numerical}
\subsection{Sensitivity of defective eigenvalue 
to perturbations}
\label{sec:sensitive}

For a matrix $M$, its spectrum (the set of eigenvalues)
is denoted by $\sigma(M)$.
It is proved that Definition \ref{thm:PS} of pseudospectra
given in Section \ref{sec:Pseudo} is equivalent with
the following definition 
\cite{RT92} \cite[Theorem 2.1]{TE05}.
\begin{df}
\label{thm:PS2}
The $\varepsilon$-pseudospectrum of a matrix
$M \in \C^{n \times n}$ with $\varepsilon >0$ is the set
of $z \in \C$ such that 
\[
z \in \sigma(M+E)
\]
for some matrix $E \in \C^{n \times n}$ with
$\|E\| < \varepsilon$.
\end{df}
\vskip 0.3cm
That is, for a given matrix, 
the $\varepsilon$-pseudospectrum is \textit{not}
the exact spectrum of the matrix,
but it is the set of exact eigenvalues of 
some perturbed matrix $M+E$ with $\|E\| < \varepsilon$.

\begin{figure}[htbp]
    \begin{tabular}{ccc}
    \hskip -2.5cm
      \begin{minipage}[t]{0.63\hsize}
       \centering
        \includegraphics[keepaspectratio, scale=0.25]{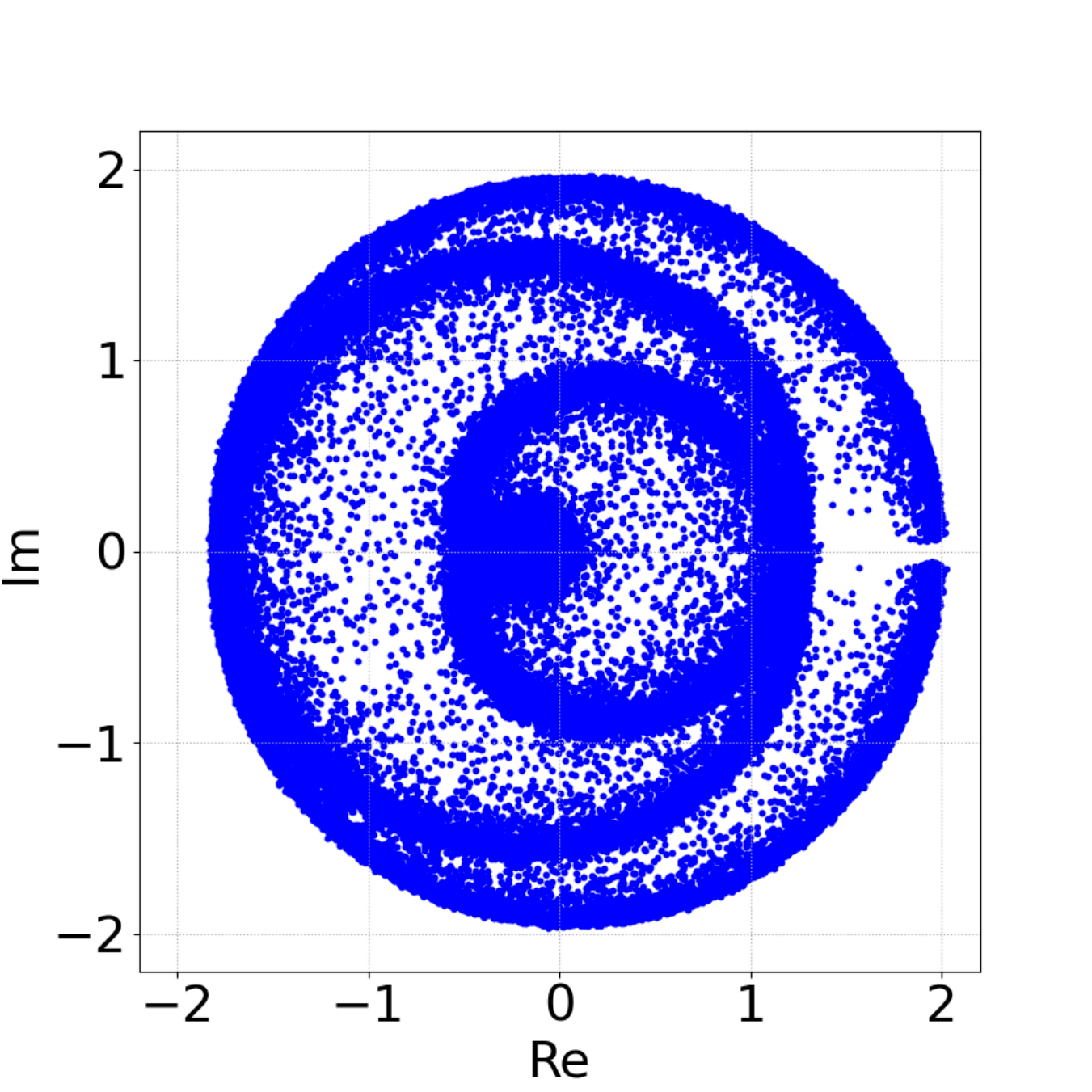}
        \subcaption{$\widetilde{\delta}=10^{-2}$}
        \label{fig:SdJdZa}
      \end{minipage} &
    \hskip -5.3cm
      \begin{minipage}[t]{0.63\hsize}
       \centering
        \includegraphics[keepaspectratio, scale=0.25]{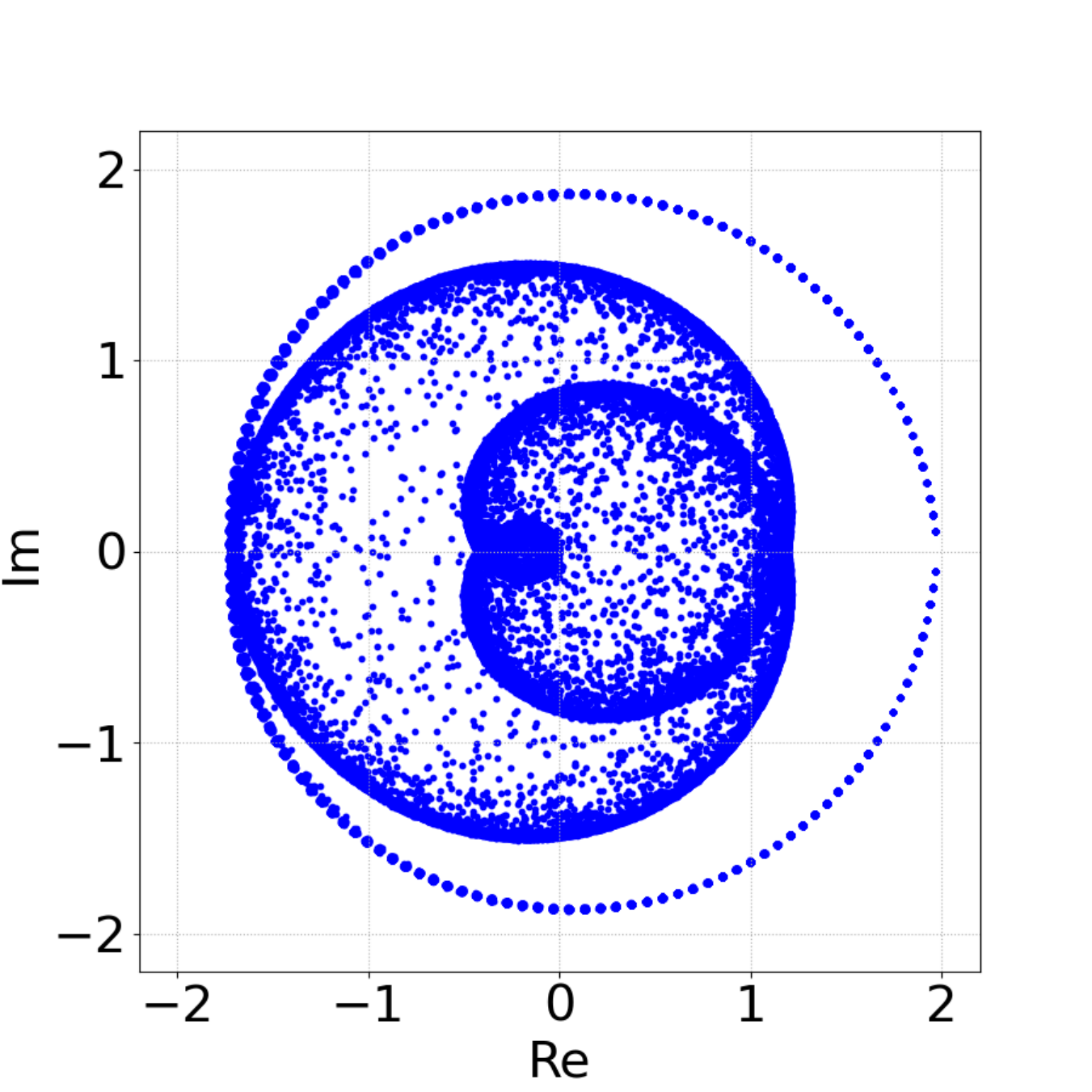}
        \subcaption{$\widetilde{\delta}=10^{-4}$}
        \label{fig:SdJdZb}
      \end{minipage} &
          \hskip -5.3cm
      \begin{minipage}[t]{0.63\hsize}
       \centering
        \includegraphics[keepaspectratio, scale=0.25]{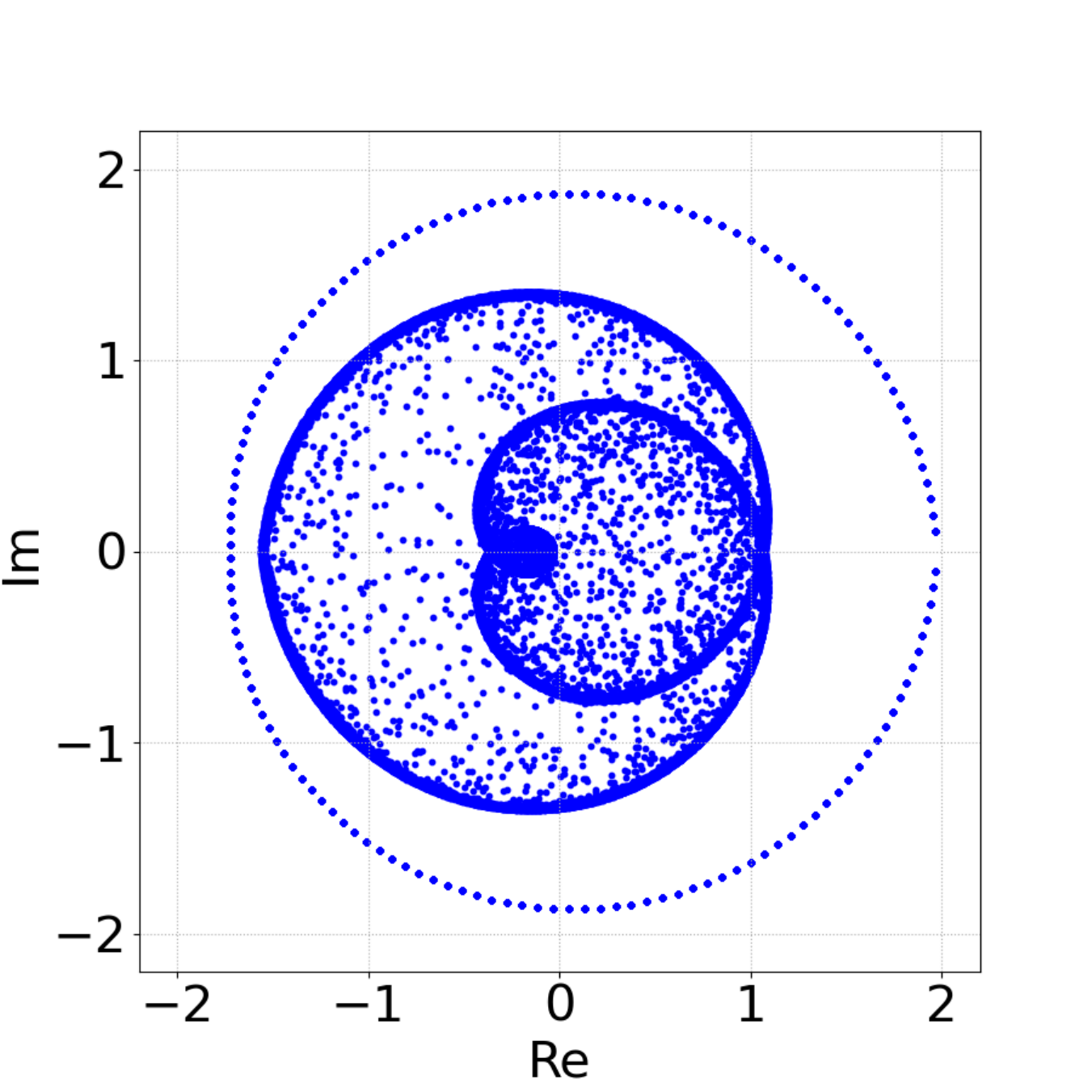}
        \subcaption{$\widetilde{\delta}=10^{-10}$}
        \label{fig:SdJdZc}
      \end{minipage} 
    \end{tabular}
     \caption{Numerically obtained eigenvalues are
     superposed for 200~i.i.d~Gaussian random perturbations
     $Z$ added to $S^{(b)}(t, \delta J)$ 
     as \eqref{eq:Gaussian1} with each value of 
     $\widetilde{\delta}$, where $n=500$, $t=2$, $b=1$, 
     and $\delta=10^{-2}$. 
     The outlier eigenvalue $\lambda_1(t)$ 
     is not drawn in each figure, since it is located 
     outside the frame.}
     \label{fig:SdJdZ}
  \end{figure}

Here we add Gaussian random perturbations
to our dynamical systems \eqref{eq:model} as
\begin{equation}
S^{(b)}(t, \delta J) + \widetilde{\delta} Z
=S^{t+1}+b S^{t+2} + \delta J 
+ \widetilde{\delta} Z,
\label{eq:Gaussian1}
\end{equation}
where the entries of $Z=(Z_{jk})_{1 \leq j, k \leq n}$ 
are given by \eqref{eq:Z} and $\widetilde{\delta} \in \C$.
Figure \ref{fig:SdJdZ} shows the plots of 
numerically obtained eigenvalues 
for $n=500$, $t=2$, $b=1$, and $\delta=10^{-2}$.
Each of Figs.~\ref{fig:SdJdZa}--\ref{fig:SdJdZc}
shows the superpositions of the numerically obtained 
eigenvalues for 200 i.i.d.~$Z$.
The following are observed.

\begin{description}
\item{(i)} \,
Fig.\ref{fig:SdJdZa} shows the numerical result 
when $\widetilde{\delta}$ is equal to 
$\delta=10^{-2}$. The obtained plots are 
similar to Fig.\ref{fig:SdZb} 
of the superpositions of 50 samples 
of numerically obtained eigenvalues for \eqref{eq:S_Gauss}
with $n=5000$, $m=t+1=3$, $b=1$,
and $\widetilde{\delta}=1/\sqrt{2n}=10^{-2}$,
although in \eqref{eq:S_Gauss} 
the all-ones matrix $J$ was not
added; in other words, $\delta=0$.
This suggests that, if $\delta=\widetilde{\delta}$,
both of $\delta J$ and $\widetilde{\delta} Z$ 
are acting as perturbations to 
$S^{t+1}+b S^{t+2}$ and hence
the eigenvalues of the 
perturbed system \eqref{eq:Gaussian1} tend to trace out
the symbol curve of the Toeplitz operator
$\widehat{S}^{t+1}+b \widehat{S}^{t+2}$, $b=1$. 
See Fig.\ref{fig:SdZc} for the symbol curve 
\cite{BS99,TE05} for $m=t+1=3$ and $b=1$. 

\item{(ii)} \,
As shown by Fig.\ref{fig:SdJdZb}, if
$\widetilde{\delta}$ is set to be much smaller than
$\delta$, then the situation is realized such that 
the system $S^{(b)}(t, \delta J)$ including 
the term $\delta J$ is perturbed by $Z$. 
Now the outmost curve represents
the $p_1(t)$ non-zero exact eigenvalues
of the unperturbed system 
solving \eqref{eq:eigenvalues} of Theorem \ref{thm:ev1}, 
except the outlier eigenvalue 
$\lambda_1(t)$ characterized by
Proposition \ref{thm:ev2}.
(See Proposition \ref{thm:ev3} asserting a circular
configurations of the non-zero exact eigenvalues 
of $S^{(b)}(t, \delta J)$ when $n$ is sufficiently large.) 
Notice that 
$S^{(b)}(t, \delta J)=S^{t+1}+b S^{t+2}
+ \delta J$, $t =1,2, \dots, T$ are still 
Toeplitz matrices, but they are neither \textit{banded} 
nor in the Wiener class 
due to the all-ones matrix $J$.
Hence we do not have any corresponding
Toeplitz operators nor symbol curves.
(For the relationship between the peudospectra
of banded Toeplitz matrices and
the exact spectra of corresponding Toeplitz
operators, see \cite{RT92}, \cite[Section 7]{TE05}.) 
When $\widetilde{\delta}=10^{-10}$ 
as shown by Fig.\ref{fig:SdJdZc},
we observe separation of the
symbol curve of
$\widehat{S}^{t+1}+b \widehat{S}^{t+2}$ with $t=2$
(see Fig.\ref{fig:SdZc})
into the outmost part and
the inner part and the size of the latter is
reduced. 
Such phenomena in pseudospectrum processes
exhibiting \textit{separation} of symbol curves
and \textit{dilatation} of their inner parts 
have not been reported in the previous studies
on banded Toeplitz matrices with random perturbations
\cite{BGKS22,BKMS21,BPZ19,BPZ20,BZ20,SV21}.

\item{(iii)} \,
The non-zero exact eigenvalues of
the original $S^{(b)}(t, \delta J)$
are \textit{insensitive} to random perturbation
and the outmost part consisting of them
keeps almost the same circular configuration.
The inner part, in which the eigenvalues
of perturbed system are densely distributed, expresses the
pseudospectrum of including $\lambda_0$
of the original system.
It continues to reduce its size  
as $\widetilde{\delta}$ decreases down to
$\widetilde{\delta}=10^{-30}$. 
Even if we put $\widetilde{\delta}=0$, however, 
we observed the very similar result with 
the pattern for $\widetilde{\delta}=10^{-30}$. 
This means that rounding errors in our personal computer
will be $\simeq 10^{-30}$ and they 
also produce perturbations to $S^{(b)}(t, \delta J)$,
and hence we can not realize $\widetilde{\delta} \to 0$
limit in the present numerical study. 
Such high \textit{sensitivity of the defective eigenvalue}
$\lambda_0$ to very weak perturbations is consistent with 
the persistency of the $\varepsilon$-pseudospectrum 
including $\lambda_0$ as $\varepsilon \to 0$
mentioned in Remark \ref{thm:remarks_PS} 
(i) in Section \ref{sec:Pseudo}. 
\end{description}

\subsection{$(t, n)$-dependence of
pseudospectra}
\label{sec:t_n}

\begin{figure}[ht]
\begin{center}
\includegraphics[width=0.90\textwidth]{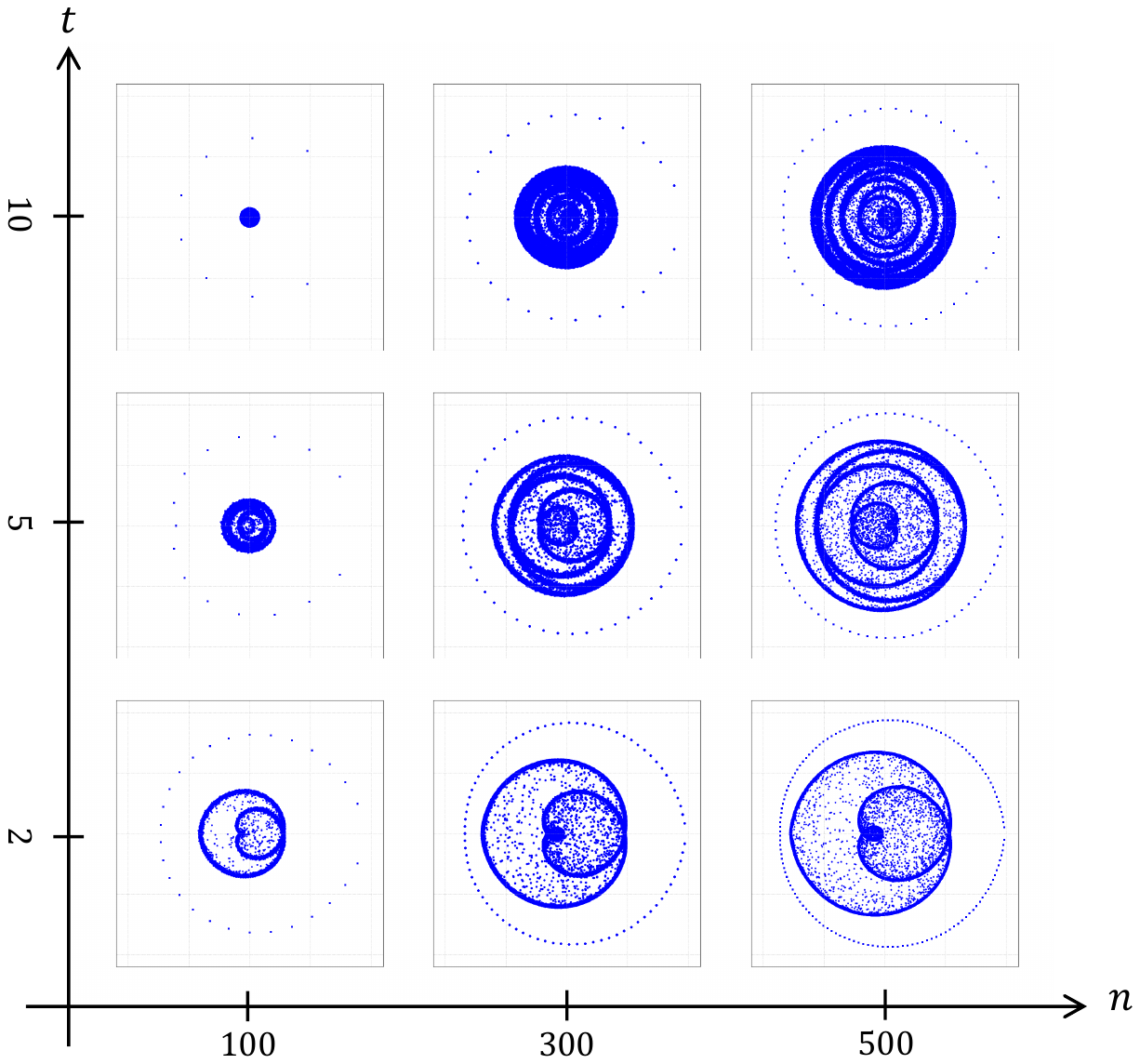}
\caption{ 
     Numerically obtained eigenvalues are
     superposed for 200~i.i.d~Gaussian random perturbations
     $Z$ added to $S^{(b)}(t, \delta J)$ 
     as \eqref{eq:Gaussian1} with 
     $b=1$, $\delta=10^{-2}$, and 
     $\widetilde{\delta}=10^{-10}$.
     Dependence on $n$ and $t$ is shown.
     The pseudospectrum including $\lambda_0$ 
     is observed as an inner domain fulfilled by dots.
     It shrinks with increment of $t$ showing
     the relaxation process of the
     defectivity of $\lambda_0$ for each $n$. 
     On the other hand, 
     as $n$ increases the inner part 
     shows expansion. The curves lined up by
     the eigenvalues of perturbed systems seem to 
     draw the inner parts of symbol curves of
     $\widehat{S}^{t+1}+b \widehat{S}^{t+2}$. 
}
\label{fig:table}
\end{center}
\end{figure}
\begin{figure}[ht]
 \centering
    \begin{tabular}{cc}
        \includegraphics[keepaspectratio, scale=0.35]
        {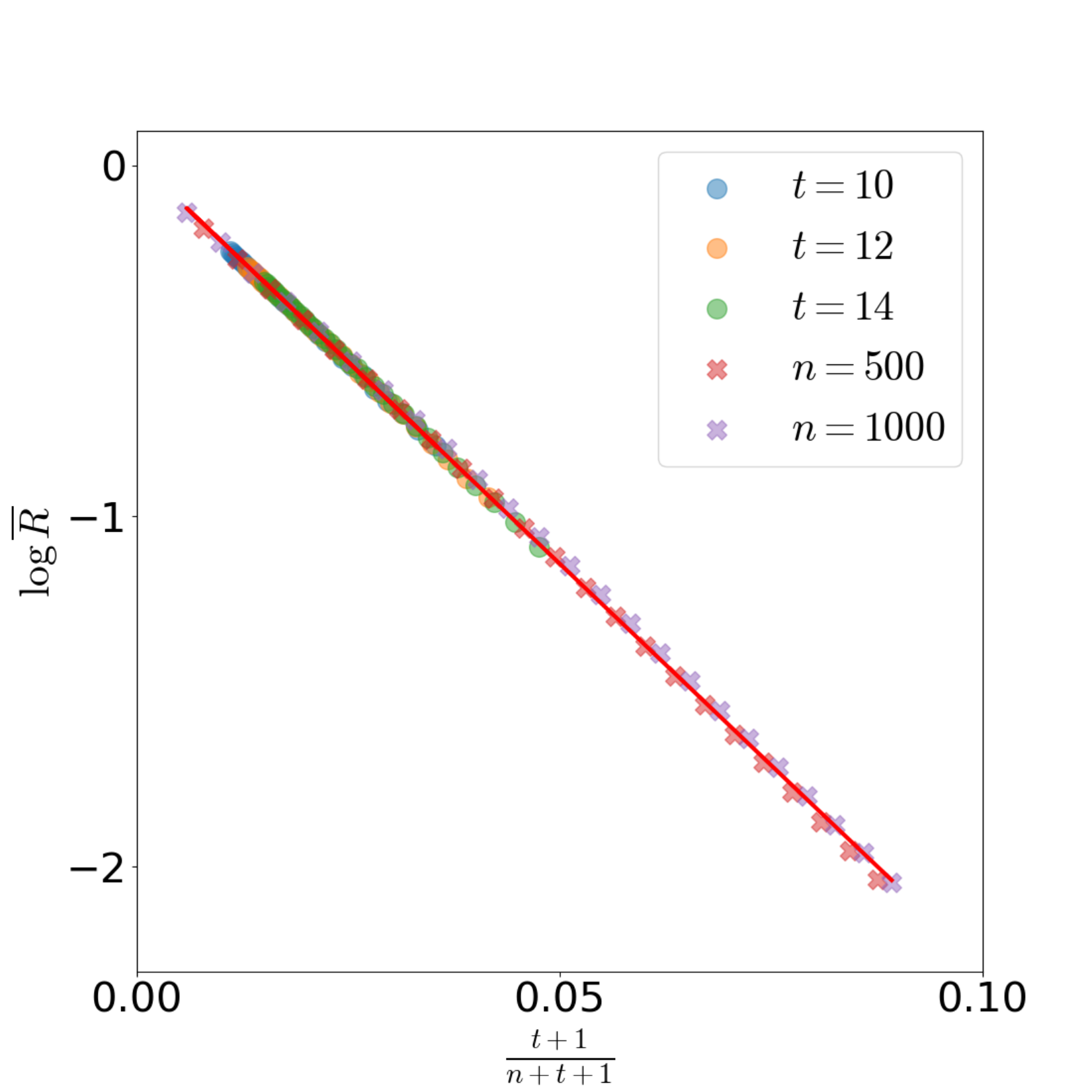}
    \end{tabular}
     \caption{Numerically evaluated radius
     $\overline{R}(t, n)$ defined by \eqref{eq:Rbar} 
     are plotted for
     the perturbed system \eqref{eq:Gaussian1} with
     $b=0$, $\delta=10^{-2}$, 
     and $\widetilde{\delta}=10^{-10}$ for
     a variety of values of $t$ and $n$.
     The results are superposed 
     for $t=8, 10, 12$ with increasing $n$ by 20 
     from 120 to 980,
     for $n=500$ with increasing $t$ by 2 from 2 to 48,
     and for $n=1000$ with increasing $t$ by 4 from 2 to 98,
     respectively. 
     The linear fitting \eqref{eq:fitting} of all these data is
     shown by the red line, which gives
     $c_1 \simeq - 23$ and $c_2 \simeq 0.02$.}
\label{fig:plot}
  \end{figure}

Theorem \ref{thm:upperPS} suggests that, 
with given $b$
if we denote the linear size of the 
$\varepsilon$-pseudospectrum including $\lambda_0$
as $R_{\varepsilon}$ for the present system
$S^{(b)}(t, \delta J)$,
it depends on $\varepsilon$, $t$, and $n$ 
in the form
\begin{equation}
\log R_{\varepsilon}(t, n) \simeq c(\varepsilon, t, n) 
\frac{t+1}{n+t+1}, \quad 
t=1,2, \dots, T:=n-2.
\label{eq:plot}
\end{equation}
Here the coefficient $c$ depends mainly on $\varepsilon$
and its dependence on $t$ and $n$ will be small only
though their logarithms.
As mentioned by Remark \ref{thm:remarks_PS},
with a fixed $0< \varepsilon \ll 1$,
$R_{\varepsilon}(t, n)$ will decrease with increment of $t$
and will expand with increment of $n$.

Figure \ref{fig:table} shows
plots of the numerically obtained eigenvalues
superposed for 200 i.i.d. Gaussian perturbed 
systems \eqref{eq:Gaussian1} with
$b=1$, $\delta=10^{-2}$, and $\widetilde{\delta}=10^{-10}$.
Here,  we set
$n=100$, $200$, $500$ and $t=2, 5, 10$.
In each figure, 
the non-zero exact eigenvalues $\lambda_j(t)$,
$j=2, 3, \cdots, p_1(t)+1$ make the outmost
circular configuration. 
Here $p_1(t)=\lfloor (n-1)/(t+1) \rfloor$,
and the radius of the circle 
increases (resp. decreases) as $n$ (resp. $t$) 
increases.
The pseudospectrum including $\lambda_0$
is observed as an inner disk fulfilled by dots,
which indeed shrinks with increment of $t$
and expands with increment of $n$, 
keeping separation from the 
outmost circle of exact eigenvalues.
Notice that as $t$ increases more complicated
structures appear in the inner disk. 
They reflect the inner parts of the symbol curves
of $\widehat{S}^{t+1}+b \widehat{S}^{t+2}$
which become more complicated as $t$ increases
\cite{MKS24}.

Given $b$, 
we performed numerical calculation of eigenvalues
of the randomly perturbed system \eqref{eq:Gaussian1} 
with $\delta=10^{-2}$
and $\widetilde{\delta}=10^{-10}$
for a variety of $t$ and $n$.
For each numerical result, which is represented
by the similar figure 
to Fig.\ref{fig:SdJdZc},
we eliminated the plots of non-zero eigenvalues
$\lambda_j(t)$, $j=1,2, \dots, p_1(t)+1$
making the outmost circle.
We write the remaining eigenvalues as
$\widetilde{\lambda}_j(t)$,
$j=1,2, \dots, m$,
where $m$ is the total number of them.
Then we numerically evaluated the mean radius 
$\overline{R}(t, n)$ of them, 
\begin{equation}
\overline{R}(t, n) :=
\frac{1}{m} \sum_{j=1}^m |\widetilde{\lambda}_j(t)|.
\label{eq:Rbar}
\end{equation}
We have observed that, 
when $b=0$,
$\overline{R}(t, n)$ is approximately
equal to the radius 
of the boundary of the region 
where $\widetilde{\lambda}_n(t)$ are distributed,
which we write $\widetilde{R}(t,n)$.
For $b \not=0$, it was numerically confirmed that
$\overline{R}(t,n) \simeq \gamma \widetilde{R}(t,n)$
with a proportionality coefficient $\gamma <1$,
which depends on $b$; 
e.g., $\gamma \simeq 2/3$ for $b=1$.
Figure \ref{fig:plot} shows 
$\log \overline{R}(t, n)$ versus
$(t+1)/(n+t+1)$ in the simple case with $b=0$.
The linear fitting 
\begin{equation}
\log \overline{R}(t, n) = c_1 \frac{t+1}{n+t+1}
+ c_2
\label{eq:fitting}
\end{equation}
works well with $c_1 \simeq - 23$ and $c_2 \simeq 0.02$.
It is expected that $c_1$ is approximately equal to
$\log \varepsilon$. 
Then the above fitting will give
$\varepsilon=e^{-23} \simeq 1.0 \times 10^{-10}$.
This is consistent with the choice of the coefficient
$\widetilde{\delta} = 10^{-10}$ 
in the numerical simulation of \eqref{eq:Gaussian1}.
Hence, although the formula \eqref{eq:plot} 
is an approximation, it seems to be valid. 
We notice that Theorem \ref{thm:upperPS}
with Remark \ref{thm:remarks_PS}
and Fig.\ref{fig:plot}
supports Conjecture 15 given in \cite{MKS24}. 

\SSC
{Future Problems}
\label{sec:future}

We list out future problems.
\begin{description}
\item{(1)} \,
For $n \geq 2$, let $h$ be a polynomial of 
${\rm deg} \, h \geq 1$, that is,
\begin{equation}
h(s)=h(s, \{b_j\}_{j \geq 1}):= \sum_{j=1}^{n-1} b_j s^j
\label{eq:hs}
\end{equation}
with $b_j \in \C$, $j=1, 2, \dots, n-1$.
We consider a nilpotent Toeplitz matrix
\begin{equation}
S_n(t, \{b_j\}_{j \geq 1})
:=S_n^{t+1} (I+ h(S_n, \{b_j \}_{j \geq 1})), 
\quad t=1,2, \dots, T:=n-2,
\label{eq:Snb}
\end{equation}
where $S_n$ is the shift matrix \eqref{eq:S} of size $n$
and $b_j \in \C$, $j \geq 1$.
At each time $t \geq 1$, we consider a one parameter
(the matrix size $n=2,3,\dots$) family, 
$\{S_n(t, \{b_j\}_{j \geq 1})\}_{n \geq 2}$.
If the Toeplitz matrices are banded; that is,
$b_j=0$ for $j \geq {^{\exists}w}$, or
in the Wiener class; $\sum_{j \geq 1} |b_j|<\infty$,
the symbol is defined by
\begin{equation}
f(z)=f(z, t, \{b_j\}_{j \geq 1}) :=z^{t+1}(1+h(z)).
\label{eq:symbol}
\end{equation}
For a given point $z \in \C \setminus f(\T)$,
$w(f, z)$ is defined to be the \textit{winding number}
of the curve $f:=f(\T)$ about $z$ 
in the usual positive (counterclockwise)
sense.
The following is proved \cite{RT92,TE05}.
\begin{description}
\item{(i)} \,
Consider the triangular Toeplitz operator
corresponding to \eqref{eq:Snb}
\[
\widehat{S}(t, \{b_j\}_{j \geq 1})
:=\widehat{S}^{t+1}(\widehat{I}
+h(\widehat{S}, \{b_j \}_{j \geq 1})),
\]
where $\widehat{I}$ is the identity operator.
Then its spectrum is given by
\begin{equation}
\sigma(\widehat{S}(t, \{b_j\}_{j \geq 1}))
= f(\T) \cup \{z \in \C: w(f, z) \not=0\};
\label{eq:spectra}
\end{equation}
that is, the symbol curve together with all the points
enclosed by the symbol curve with nonzero 
winding numbers. 
\item{(ii)} \,
At any point $z \in \sigma(\widehat{S}(t, \{b_j\}_{j \geq 1}))$,
for some $C>1$ and for all sufficiently large $n$,
\begin{equation}
\| (z I - S_n(t, \{b_j\}_{j \geq 1})^{-1} \|
\geq C^n.
\label{eq:Toeplitz}
\end{equation}
\end{description}
Comparing this fact with Definitions \ref{thm:PS}
and \ref{thm:PS2}, we can conclude that
if we add dense random perturbations to
$S_n(t, \{b_j\}_{j \geq 1})$ at each time
$t=1,2, \dots, T$, 
the eigenvalues tends to fill the spectra
$\sigma(\widehat{S}(t, \{b_j\}_{j \geq 1})$.
Moreover, if we plot the numerically obtained 
eigenvalues for the perturbed system,
the symbol curve $f(\T)$ seems to be lined up
by the dots, as demonstrated by
Fig.\ref{fig:SdZ}. 
In this sense, the pseudospectra of
$S_n(t, \{b_j \}_{j \geq 1})$, $n \geq 2$, $t=1,2, \dots, T$
are well characterized by the spectra
of the corresponding triangular Toeplitz operators
$\widehat{S}(t, \{b_j \}_{j \geq 1})$, $t=1,2, \dots, T$
and their symbol curves.
On the other hand, our matrix
$S_n^{(b)}(t, \delta J)$, 
for $n \geq 2$, $t=1,2, \dots, T$
given by \eqref{eq:model} is not banded
due to addition of the all-ones matrix $J$.
Hence there is no corresponding 
Toeplitz operator nor symbol curve.
Nevertheless, we have observed the phenomenon
such that if we add the complex random Gaussian
matrices as perturbations,
the numerically obtained eigenvalues are
lined up along some curves and they tend to
fill only inside of the domains
where that curve exists.
As demonstrated by Figs.\ref{fig:SdJdZ} and \ref{fig:table},
these curves seem to be the
\textit{inner parts} of the symbol curve $f(\T)$
of $\widehat{S}^{t+1}+ b \widehat{S}^{t+2}$,
which is obtained by eliminating the
outmost closed simple curve
and is composed of $t-1$ smaller closed
simple curves osculating each other.
(See also Section 4.2 in \cite{MKS24}.)
The outmost closed simple curve
consists of the non-zero eigenvalues of
the unperturbed matrix and
they are insensitive to perturbations.
As shown by Figs.\ref{fig:SdJdZ} and \ref{fig:table},
the inner parts are shrinking in time 
and hence the gaps between the outmost curve
consisting of the original eigenvalues and
the inner part fulfilled by eigenvalues of perturbed
systems become larger as time $t$ is passing.
Such phenomena showing 
separation of symbol curves and
size-reduction of their inner parts 
have not been reported in the previous studies of
\textit{banded} Toeplitz matrices with random perturbations
(see \cite{BGKS22,BKMS21,BPZ19,BPZ20,BZ20,SV21} 
and references therein). 
The present new phenomena 
found in the pseudospectrum processes
express the relaxation
processes of the defective eigenvalue $\lambda_0$.
Mathematical understanding of such separation and 
dilatation of symbol curves will be 
a challenging future problem.

\begin{figure}[htbp]
    \begin{tabular}{ccc}
    \hskip -2.5cm
      \begin{minipage}[t]{0.63\hsize}
       \centering
        \includegraphics[keepaspectratio, scale=0.25]{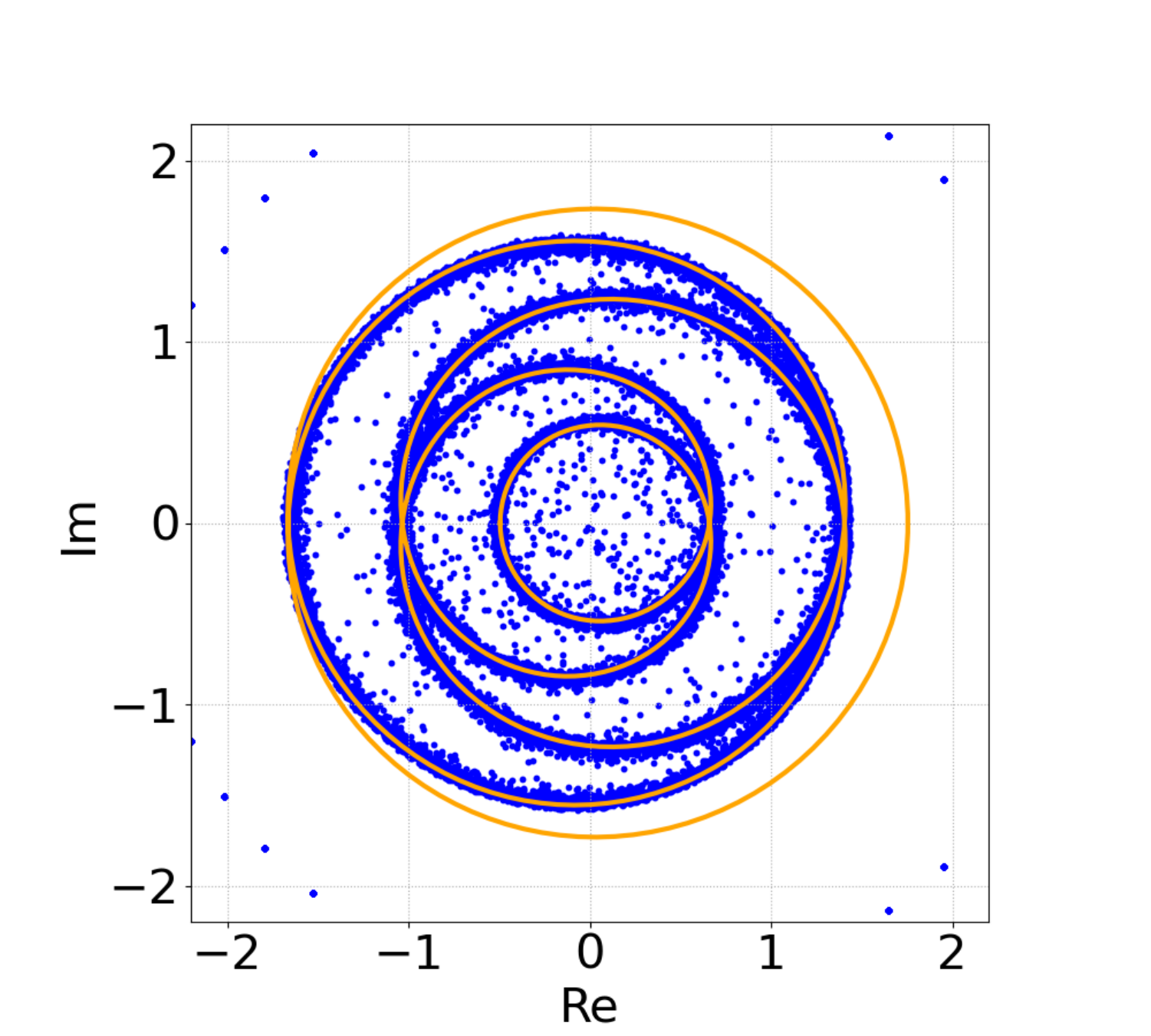}
        \subcaption{$b_1=2, b_j=0 \, (j\geq 2)$}
        \label{fig:reduceA}
      \end{minipage} &
    \hskip -5.3cm
      \begin{minipage}[t]{0.63\hsize}
       \centering
        \includegraphics[keepaspectratio, scale=0.25]{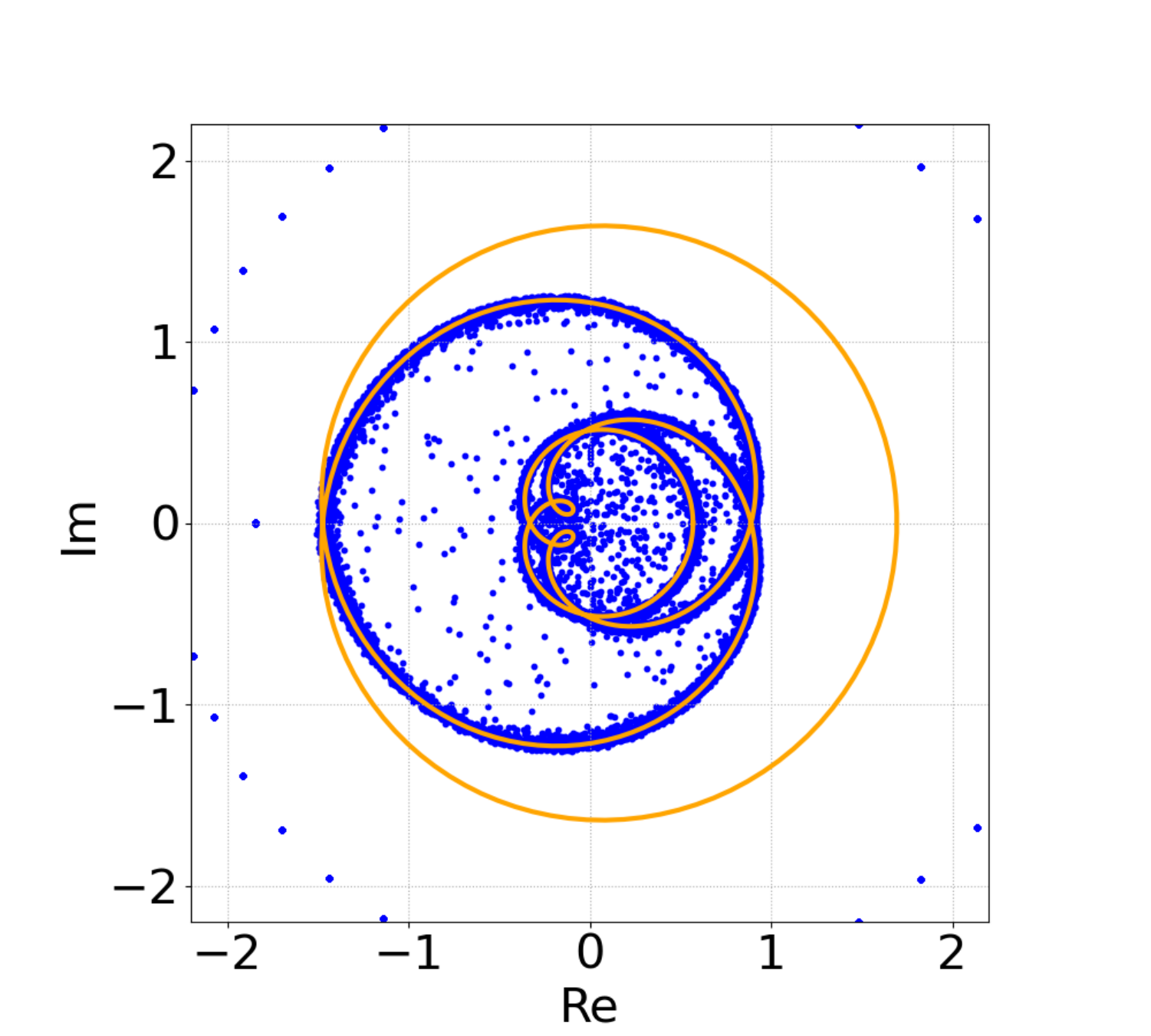}
        \subcaption{$b_1=b_2=1, b_j=0 \, (j\geq 3)$}
        \label{fig:reduceB}
      \end{minipage} &
          \hskip -5.3cm
      \begin{minipage}[t]{0.63\hsize}
       \centering
        \includegraphics[keepaspectratio, scale=0.25]{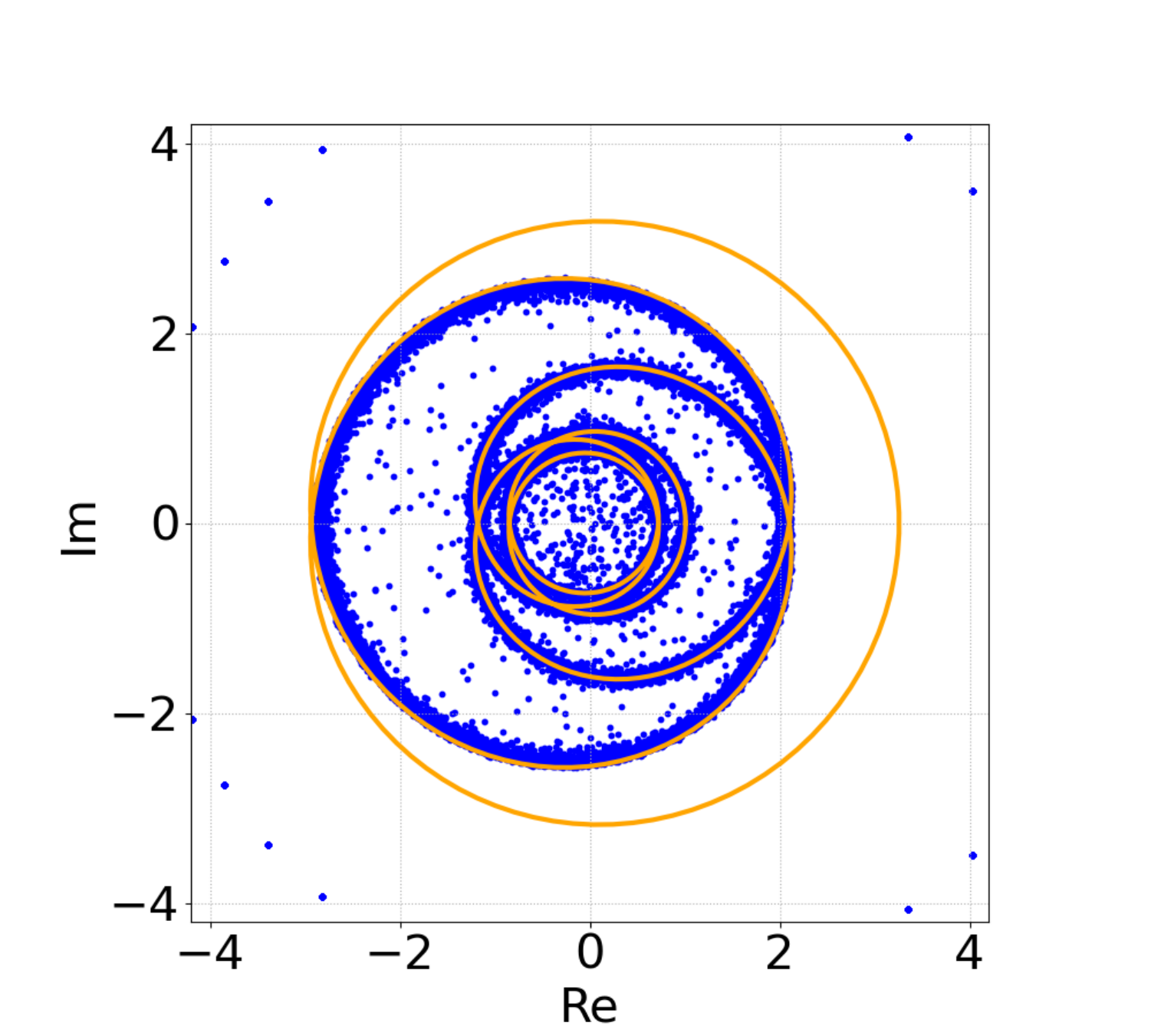}
        \subcaption{$b_1=2, b_2=3, b_j=0 \, (j\geq 3)$}
        \label{fig:reduceC}
      \end{minipage} 
     \end{tabular}
     \caption{Numerically obtained eigenvalues are
     superposed for 200~i.i.d~Gaussian random perturbations
     $Z$ added to $S_n(t, \{b_j\}_{j \geq 1})+\delta J$ 
     as \eqref{eq:reduction}
     with the specified choices of $\{b_j\}_{j \geq 1}$, 
     where $n=200$, $t=3$, 
     $\delta=10^{-2}$, and $\varepsilon=10^{-10}$. 
     The symbol curves with size reduction \eqref{eq:f_epsilon}
     are drawn by red curves.
     In the present cases, $\varpi=(3+1) \times \{200/(3+1)\}=200$
     and hence 
     $\varepsilon^{1/\varpi}=(10^{-10})^{1/200}=10^{-1/20}
     =0.891 \cdots$.
     }
     \label{fig:reduction}
  \end{figure}
  
Based on numerical study, we propose 
a conjecture as explained below.
We put 
\[
\varpi=\varpi(t, n) := (t+1) k_0(t),
\quad t=1,2, \dots, T, 
\]
where $k_0(t)$ is the index of $\lambda_0$
given by \eqref{eq:k02}. 
Assume that $\varepsilon >0$ is small enough so that
the $\varepsilon$-pseudospectra are separated
without any intersection into one 
$\varepsilon$-pseudospectrum including $\lambda_0$,
which is denoted by $\sigma_{\varepsilon}^0$,
and other parts including $\lambda_j(t)$,
$j=1,2, \dots,$ $p_1(t)+1$. 
Consider the symbol curve with size reduction
\begin{equation}
f_{\varepsilon}
:=\{f(r e^{i \theta}): r=\varepsilon^{1/\varpi},
\theta \in [0, 2\pi)\}, 
\label{eq:f_epsilon}
\end{equation}
where $f$ is given by \eqref{eq:symbol}.
Let $\cI =\cI(t, \{b_j\}_{j \geq 1}) \subset
[0, 2 \pi)$ be the interval
of the argument $\theta$ such that 
$\{f(\varepsilon^{1/\varpi} e^{i \theta}): \theta \in \cI\}$ 
provides the outmost closed simple curve of $f_{\varepsilon}$.
That is, if we define
$\theta_0:= \min\{ \theta \in (0, 2 \pi) :
{\rm Im} f(\varepsilon^{1/\varpi} e^{i \theta})=0\}$,
(i.e., the smallest positive $\theta$
such that the curve \eqref{eq:f_epsilon}
intersects with the real axis), then
$\cI=[0, \theta_0) \cup [2 \pi-\theta_0, 2 \pi)$.
We conjecture that 
\begin{equation}
\sigma_{\varepsilon}^0(
S_n(t, \{b_j\}_{j \geq 1})+ \delta J)
=\widetilde{f}_{\varepsilon}
\cup \{z \in \C: w(\widetilde{f}_{\varepsilon}, z)
\not=0\},
\label{eq:conjecture1}
\end{equation}
where 
\begin{equation}
\widetilde{f}_{\varepsilon}
:=\{f(r e^{i \theta}): r=\varepsilon^{1/\varpi},
\theta \in [0, 2\pi) \setminus \cI\},
\label{eq:tilde_f}
\end{equation}
and $w(\widetilde{f}_{\varepsilon}, z)$
denotes the winding number of the curve 
$\widetilde{f}_{\varepsilon}$ about $z$.
Notice that \eqref{eq:conjecture1} does not
depend on $\delta$. 
We have numerically studied the 
randomly perturbed systems,
\begin{equation}
S_n(t, \{b_j\}_{j \geq 1})+ \delta J +
\varepsilon Z, \quad t=1,2, \dots, T, 
\label{eq:reduction}
\end{equation}
where $Z=(Z_{jk})_{1 \leq j, k}$
is the Gaussian random matrix with
the entries \eqref{eq:Z} and
$0< \varepsilon \ll \delta \ll 1$. 
Figure \ref{fig:reduction} shows three examples of
the superpositions of numerically obtained
eigenvalues. 
There the symbol curves
with size reduction \eqref{eq:f_epsilon} 
are also drawn by red curves.
The eigenvalues of the randomly perturbed systems
tend to trace out 
the inner parts \eqref{eq:tilde_f} of the size-reduced
symbol curves \eqref{eq:f_epsilon}. 
Independence of 
$\sigma_{\varepsilon}^0(S_n(t, \{b_j\}_{j \geq 1})+\delta J)$ 
on $\delta$ has been numerically confirmed 
provided $\delta \gg \varepsilon$.

\item{(2)} \,
We have established a procedure to 
determine the generalized eigenspace
associated with $\lambda_0$ in Section \ref{sec:generalES}
without using any information of
the eigenspaces associated with the non-zero
eigenvalues.
The explicit expressions of the generalized
eigenvectors for $\lambda_0$ are
given in Appendix \ref{sec:b0} 
in the case $b=0$.
Similar calculation is desired for
the general case with $b \not=0$.
In particular, explicit calculation of the generalized 
condition numbers $\kappa_0^{(\ell)}(t)$ 
is an important open problem.

For the systems \eqref{eq:Snb}, 
here we have reported our detail study 
in the special case with $b_1=b \in \C$
and $b_j=0$, $j \geq 2$.
Generalization including $b_j$, $j \geq 2$
should be studied.
See Appendix \ref{sec:general}.

Another direction of generalization of the present
dynamics will be given by replacing the all-ones
matrix $J$ by a matrix in the form, 
$M=(\m \, \, c_2 \m \, \, c_3 \m \, \, \cdots \, \, 
c_n \m)^{\sT}$,
where $\m \in V_n$ and $c_j \in \C$, $j=2,3, \dots, n$.
In this case, 
$M \c=\bra \v, \overline{\m} \ket \c$
with $\c=(1, c_2, c_3, \dots, c_n)^{\sT}$
and hence 
$\1$ and $\alpha(\v(t))=\bra \v(t), \1 \ket
=\sum_{j=1}^n v_j(t)$ shall be replaced by $\c$ and
$\bra \v(t), \overline{\m} \ket$, respectively, 
in calculations.
It will be possible to consider the 
discrete-time stochastic processes,
in which the vectors 
$\m$ and $\c$ are randomly distributed
following some probability laws.
(See \cite{PS17} for the random nilpotent matrix model.)

We expect a meaningful connection between
the present calculations of
our dynamics of nilpotent Toeplitz matrices
and the representation theory of 
the semisimple Lie groups via time-evolutionary
Young diagrams \cite{CM93,Gan81,PS17} demonstrated by
Fig.\ref{fig:Young}.

\item{(3)} \, As mentioned in Section \ref{sec:introduction},
one of the motivations of the present study
is the observation by Burda et al.~\cite{Burda15}
of the non-Hermitian matrix-valued BM
started from a nonnormal and defective matrix $S$.
In the non-Hermitian matrix-valued stochastic
processes,
the coupling between the eigenvalue processes
and the time evolution of 
the \textit{eigenvector-overlap matrices}
is essential \cite{BCH24,BD20,Burda15,EKY23,GW18}.
The square roots of the 
diagonal elements of the eigenvector-overlap
matrices are the condition numbers of
eigenvalues.
Recently, statistical properties of the 
\textit{conditional numbers} have been extensively
studied \cite{BGZ18,BCH24,BD20,CEX26}. 
In \eqref{eq:generalCN} in Section \ref{sec:Jordan},
we introduced a notion of 
\textit{generalized condition numbers},
which represent
overlap between the left- and the 
right-eigenvectors in the same Jordan block
associated with the defective eigenvalue $\lambda_0$.
It will be a challenging problem to introduce standard
statistical ensembles and stochastic processes
of random matrices in the Jordan canonical form
and clarify any universal probability laws of 
defective eigenvalues
and generalized condition numbers. 
We hope that the present study of the deterministic
processes will lead us further understanding
of the coupling systems of
eigenvalue processes and eigenvector-overlap
processes in nonnormal matrix-valued 
stochastic processes. 
\end{description}

\vskip 1cm
\noindent{\bf Acknowledgements} \,
A part of this study was presented by MK
in the conference 
`Random Matrices and Related Topics in Jeju',
May 6--10, 2024, 
held in Jeju Island, Korea.
MK and TS would like to thank
Sung-Soo Byun, 
Nam-Gyu Kang, 
and 
Kyeongsik Nam 
very much for organizing such a wonderful conference.
MS was supported by JSPS KAKENHI Grant Numbers 
JP19K03674, 
JP21H04432,
JP22H05105,
JP23K25774,
and
JP24K06888.
TS was supported by JSPS KAKENHI Grant Numbers 
JP20K20884,
JP22H05105 
and 
JP23K25774. 
TS was also supported in part by JSPS KAKENHI Grant Numbers 
JP21H04432
and
JP24KK0060.
This work was also supported in part 
by the Research Institute 
for Mathematical Sciences,
an International Joint Usage/Research Center 
located in Kyoto University.

\appendix
\SSC{Proof of \eqref{eq:v0_ell2} in a general setting}
\label{sec:general}

For the shift matrix $S=S_n$ given by \eqref{eq:S}, 
we consider the dynamics, 
$(S_n(t, \{b_j\}_{j \geq 1}))_{t=1}^{T}$, defined by
\eqref{eq:Snb} with \eqref{eq:hs}. 
Note that there is a polynomial $\hhat(s)$ of 
$\deg \hhat \ge 1$ such that 
$(I+h(S))(I+\hhat(S)) = I$, since $S$ is nilpotent.  
Given $\v$, 
we consider the following problem, 
\[
 \{(S^{t+1} (I+h(S)) + \delta J\} \w = \la \w + \v.  
\] 
We focus on the case where $\la=\la_0 =0$, i.e., 
\begin{equation}
 \{(S^{t+1} (I+h(S)) + \delta J\} \w = \v.  
\label{eq:eq0}
\end{equation} 
By \eqref{eq:Jv=a1}, this is equivalent to 
\begin{equation}
 S^{t+1} (I+h(S)) \w = \v -\delta \alpha(\w) \trivial. 
\label{eq:eq1} 
\end{equation}
\begin{lem}
\label{thm:wsol}
Suppose $\v=(v_1,v_2,\dots,v_n)$ such that $v_j=0$ for some
 $j \ge n-t$. Then, 
\begin{equation}
\w = - \alpha\big( S^{-(t+1)} (I+\hhat(S)) \v \big) \e_1 
+ S^{-(t+1)} (I+\hhat(S)) \v 
\label{eq:solw}
\end{equation}
is a solution to \eqref{eq:eq0}, 
where $S^{-(t+1)}$ is defined by \eqref{eq:S_inverse}.
\end{lem}
\noindent{\it Proof} \,
By comparing the $j$-th coordinate of both sides of \eqref{eq:eq1}, 
we see that $\alpha(\w)=0$ in \eqref{eq:eq1}, which is equivalent to 
\begin{equation}
 S^{t+1} \w = (I+\hhat(S))\v, \quad \alpha(\w)=0. 
\label{eq:eq2} 
\end{equation}
Therefore, for any $\u \in V_{t+1}$, 
\begin{equation}
\w = \u + S^{-(t+1)} (I+\hhat(S)) \v
\end{equation}
is a solution to \eqref{eq:eq1} whenever $\alpha(\w)=0$. 
Since 
\begin{equation}
\u = - \alpha\big( S^{-(t+1)} (I+\hhat(S)) \v \big) \e_1 \in
 V_1
\end{equation}
and $\alpha(\w) = 0$, we see that $\w$ given in \eqref{eq:solw} 
is a solution to \eqref{eq:eq1}. 
\qed
\vskip 0.3cm
\begin{cor}
\label{thm:v_corr}
Let $1+\hhat(s) = \sum_{k=0}^{n-1} b_k s^k$ with $b_0=1$. 
Then, for $t+2+\ell \le n$, 
\begin{equation}
\w
= \Big(\sum_{k=1}^{\ell} b_k\Big) \e_1 + 
(1-b_{\ell})\e_{t+2} - \sum_{p=1}^{\ell} b_{\ell-p}
 \e_{t+2+p}. 
\label{eq:Appendix_w}
\end{equation}
is a solution to \eqref{eq:eq0} for $\v=\v^{(\ell,1)}(t) =
 \e_1 - \e_{\ell+1}$. 
\end{cor} 
\noindent{\it Proof} \,
We consider the case where $\v = \e_1 - \e_{\ell+1}$. 
\begin{align*}
S^{-(t+1)} (I+\hhat(S)) \v 
&= S^{-(t+1)} \left(\v + \sum_{k=1}^{n-1} b_k S^k \v \right) 
\\
&= S^{-(t+1)} \v - S^{-(t+1)} \sum_{k=1}^{\ell} b_k S^k \e_{\ell+1} \\
&= \e_{t+2} - \sum_{k=0}^{\ell} b_k \e_{t+2+\ell-k} \\
&= (1-b_{\ell})\e_{t+2} - \sum_{p=1}^{\ell} b_{\ell-p} \e_{t+2+p} 
\end{align*}
for $t+2+\ell \le n$. Then,  
\[
 \alpha(S^{-(t+1)} (I+\hhat(S)) \v) = - \sum_{k=1}^{\ell}
 b_k.  
\]
From Lemma~\ref{thm:wsol}, 
a solution $\w$ to \eqref{eq:eq1} is given by 
\eqref{eq:Appendix_w}. 
The proof is complete. \qed
\begin{ex} 
Suppose $h(s) = bs$. 
Then, $\hhat(s) = \sum_{k=1}^{n-1} (-b)^k s^k$. 
For $t+2+\ell \le n$, 
\begin{align*}
\w 
&= 
\Big\{ \sum_{k=1}^{\ell} (-b)^k \Big\} \e_1 + 
\{1-(-b)^{\ell} \}\e_{t+2} 
- \sum_{p=1}^{\ell} (-b)^{\ell-p} \e_{t+2+p} \\
&= \frac{b\{-1+(-b)^{\ell}\}}{1+b} \e_1
+ 
\{1-(-b)^{\ell} \} \e_{t+2} 
- \sum_{p=1}^{\ell} (-b)^{\ell-p} \e_{t+2+p}. 
\label{eq:example}
\end{align*}
This proves \eqref{eq:v0_ell2}.
\end{ex}

\SSC{Explicit expressions of the generalized
eigenvectors associated with $\lambda_0=0$
when $b=0$}
\label{sec:b0}
Here we consider the case $b=0$. 
Remind Proposition \ref{thm:k0}. 
\begin{description}
\item{(i)} \,
The generalized right-eigenvectors for $\lambda_0=0$
are given by
\[
\v_0^{(\ell, q)}(t)
= \Big( \underbrace{0, \cdots, 0}_{(t+1)(q-1)}, 1,
\underbrace{0, \cdots, 0}_{\ell-1}, -1,
\underbrace{0, \cdots, 0}_{t-\ell+n-(t+1) q} \Big)^{\sT},
\quad q=1, 2, \dots, d_0^{(\ell)}(t).
\]

\item{(ii)} \,
When $\xi=\xi(t,n) \in \{0,1\}$, 
the left eigenvectors are given by
\[
\w_0^{(\ell, d_0^{(\ell)}(t))}(t)
=\frac{1}{t+1} \Big(
\underbrace{0, \cdots, 0}_{n-t-1},
\underbrace{1, \cdots, 1}_{\ell},
-t,
\underbrace{1, \cdots, 1}_{t-\ell}
\Big)^{\sT},
\quad
\ell=1,2, \dots, t.
\]
The generalized left-eigenvectors are then
given by
\[
\w_0^{(\ell, d_0^{(\ell)}(t)-q)}(t)
=\frac{1}{t+1} \Big(
\underbrace{0, \cdots, 0}_{n-(q+1)(t+1)},
\underbrace{1, \cdots, 1}_{\ell},
-t,
\underbrace{1, \cdots, 1}_{t-\ell},
\underbrace{0, \cdots, 0}_{q (t+1)}
\Big)^{\sT},
\]
$\ell=1,2, \dots, t$, $q=1, \dots, d_0^{(\ell)}(t)-1$.

\item{(iii)} \,
When $\xi=\xi(t,n) \in \{2,3,\dots,t\}$, 
the left-eigenvectors are determined as
\begin{equation}
\w_0^{(\ell, d_0^{(\ell)}(t))}(t)
=\frac{1}{\xi} \Big(
\underbrace{0, \cdots, 0}_{n-\xi},
\underbrace{1, \cdots, 1}_{\ell},
1-\xi,
\underbrace{1, \cdots, 1}_{\xi-\ell-1}
\Big)^{\sT},
\quad \mbox{for $\ell=1,2, \dots, \xi-1$}, 
\label{eq:w0_eigenvector2}
\end{equation}
and
\[
\w_0^{(\ell, d_0^{(\ell)}(t))}(t)
=\frac{1}{\xi} \Big(
\underbrace{0, \cdots, 0}_{n-t-1+\ell-\xi},
-\xi,
\underbrace{0, \cdots, 0}_{t-\ell},
\underbrace{1, \cdots, 1}_{\xi}
\Big)^{\sT},
\quad
\mbox{for $\ell=\xi, \xi+1, \cdots, t$}.
\]
The generalized left-eigenvectors are then
given as follows. 
Let
\[
\Omega_{t, \xi}:=-\frac{t-\xi+1}{\xi}.
\]
For $\ell=1,2, \dots, \xi-1$,
\begin{align*}
\w_0^{(\ell, d_0^{(\ell)}(t)-q)}(t)
&= \frac{1}{\xi} \Big( 
\underbrace{0, \cdots, 0,}_{(t+1)\{ d_0^{(\ell)}(t)-(q+1) \}}
\underbrace{1, \cdots, 1}_{\ell}, \quad
1-\xi, \quad
\underbrace{1, \cdots, 1}_{t-\ell},
\nonumber\\
& \qquad 
\underbrace{ \Omega_{t, \xi}, \cdots, \Omega_{t, \xi}}_{t+1}, 
\underbrace{ \Omega_{t, \xi}^2, \cdots, 
\Omega_{t, \xi}^2}_{t+1},
\cdots,
\underbrace{ 
\Omega_{t, \xi}^{q-1}, \cdots, \Omega_{t, \xi}^{q-1}}_{t+1},
\underbrace{ \Omega_{t, \xi}^q, \cdots, 
\Omega_{t, \xi}^q}_{\xi}
\Big)^{\sT},
\end{align*}
$q=1, \dots, d_0^{(\ell)}(t)-1$. 
For $\ell=\xi, \xi+1, \dots, t$, 
\begin{align*}
\w_0^{(\ell, d_0^{(\ell)}(t)-q)}(t)
&= \frac{1}{\xi} \Big( 
\underbrace{0, \cdots, 0,}_{(t+1)\{ d_0^{(\ell)}(t)-(q+1)\}+\ell}
\quad -\xi, \quad \underbrace{0, \cdots, 0}_{t-\ell},
\underbrace{1, \cdots, 1}_{t+1}, 
\nonumber\\
& \qquad 
\underbrace{ \Omega_{t, \xi}, \cdots, \Omega_{t, \xi}}_{t+1}, 
\underbrace{ \Omega_{t, \xi}^2, \cdots, 
\Omega_{t, \xi}^2}_{t+1},
\cdots,
\underbrace{ 
\Omega_{t, \xi}^{q-1}, \cdots, \Omega_{t, \xi}^{q-1}}_{t+1},
\underbrace{ \Omega_{t, \xi}^q, \cdots, 
\Omega_{t, \xi}^q}_{\xi}
\Big)^{\sT},
\end{align*}
$q=1, \dots, d_0^{(\ell)}(t)-1$.
\end{description}
Notice that $|\Omega_{t, \xi}| > 1$ if $1 \leq \xi < (t+1)/2$,
and $|\Omega_{t, \xi}| < 1$ if $(t+1)/2 < \xi \leq t$.
The exponential change by unit $t+1$ of the amplitude 
of $w_{0 j}^{(\ell, d_0^{(\ell)}(t)-q)}(t)$ 
in $j \in \{1,2, \dots, n\}$ shows that 
the generalized left-eigenvectors represent
\textit{boundary pseudomodes} associated with
the pseudospectrum including $\lambda_0$
\cite[Section 7]{TE05}.

By \eqref{eq:v0_ell1}, 
$\|\v_0^{(\ell,1)}(t) \|=\sqrt{2}$ independently of 
$\ell$.
And in the case 
$d_0^{(\ell)}(t)=\lfloor n/(t+1) \rfloor +1$, 
$\| \w_0^{(\ell, d_0^{(\ell)}(t))}(t) \|
=\sqrt{(\xi-1)/\xi} \leq \sqrt{(t-1)/t}$ 
independently of $\ell$
by \eqref{eq:w0_eigenvector2}. 
Therefore, for
$d_0^{(\ell)}(t)=k_0(t)=\lfloor n/(t+1) \rfloor +1$, 
$\kappa_0(t)=\sqrt{2(t-1)/t}$, 
and hence
\begin{equation}
t\kappa_0(t) = \sqrt{2 t(t-1)}. 
\label{eq:tkappa0_b0}
\end{equation}


\end{document}